\documentclass[final,authoryear,12p]{elsarticle} 

\usepackage{amsmath}
\usepackage{amsfonts}
\usepackage{amssymb}
\usepackage{graphicx}
\usepackage[margin=0.5in]{geometry}
\usepackage{float}
\usepackage{subcaption}
\usepackage{upgreek}
\usepackage{commath}
 \usepackage{natbib}
 
\usepackage{setspace}
\usepackage[ruled,vlined]{algorithm2e}

\usepackage{url}
\usepackage{listings}
\usepackage{xcolor}
\usepackage{lipsum}
\definecolor{codegreen}{rgb}{0,0.6,0}
\definecolor{codegray}{rgb}{0.5,0.5,0.5}
\definecolor{codepurple}{rgb}{0.58,0,0.82}
\definecolor{backcolour}{rgb}{0.95,0.95,0.92}

\lstdefinestyle{mystyle}{
	commentstyle=\color{codegreen},
	keywordstyle=\bfseries\color{green!40!black},
	numberstyle=\tiny\color{codegray},
	stringstyle=\color{codepurple},
	basicstyle=\ttfamily\footnotesize,
	breakatwhitespace=false,         
	breaklines=true,                 
	captionpos=b,                    
	keepspaces=true,                 
	numbers=left,                    
	numbersep=5pt,                  
	showspaces=false,                
	showstringspaces=false,
	showtabs=false,                  
	tabsize=2
}

\begin{document}
	
\begin{frontmatter}
\title{Towards real time assessment of earthfill dams via Model Order Reduction }

\author[1,2,3]{Christina Nasika \corref{cor1}}
 \ead{christina.nasika1@upc.edu}

\author[2,3]{Pedro Díez}
\ead{pedro.diez@upc.edu}
\author[1]{Pierre Gerard}
\ead{piergera@ulb.ac.be}
\author[1]{Thierry J. Massart}
\ead{thmassar@ulb.ac.be}
\author[2,3]{Sergio Zlotnik}
\ead{sergio.zlotnik@upc.edu}

 \address[1]{BATir, Université Libre de Bruxelles (ULB) , Avenue F.D. Roosevelt 50, B-1050 Brussels, Belgium}
 \address[2]{LaCàN, Universitat Politècnica de Catalunya, Campus Nord UPC, E-08034 Barcelona, Spain}
  \address[3]{International Centre for Numerical Methods in Engineering, CIMNE, Barcelona, Spain}
  
\cortext[cor1]{Corresponding author}

		\begin{abstract}
			
			    The use of Internet of Things (IoT) technologies is becoming a preferred solution for the assessment of tailings dams’ safety. Real-time sensor monitoring proves to be a key tool for reducing the risk related to these ever-evolving earth-fill structures, that exhibit a high rate of sudden and  hazardous failures.   \par
			In order to optimally exploit real-time embankment monitoring, one major hindrance has to be overcome: the creation of a supporting numerical model for stability analysis, with rapid-enough response to perform data assimilation in real time. A model should be built, such that its response can be obtained faster than the physical evolution of the analyzed phenomenon. In this work, Reduced Order Modelling (ROM) is used to boost computational efficiency in solving the coupled hydro-mechanical system of equations governing the problem.  \par
			The Reduced Basis method is applied to the coupled hydro-mechanical equations that govern the groundwater flow, that are made non-linear as a result of considering an unsaturated soil. The resulting model’s performance is assessed by solving a 2D and a 3D problem relevant to tailings dams’ safety. The ROM technique achieves a speedup of 3 to 15 times with respect to the full-order model (FOM) while maintaining high levels of accuracy.

		\end{abstract}

\begin{keyword}
	Model Order Reduction \sep Reduced Basis Method \sep  Hydro-mechanical \sep Coupled problem \sep  Non-linear problems \sep  Tailings dam
\end{keyword}

\end{frontmatter}

	\section{Introduction}
Tailings is a common by-product of the process of extracting valuable minerals and metals from mined ore.  They usually take the form of a liquid slurry made of fine mineral particles, created as mined ore is crushed, ground and processed. The volume of tailings is normally far in excess of the liberated resource and the tailings often contain potentially hazardous contaminants. It is usual practice for tailings to be stored in isolated impoundments under water and behind dams \citep{kossoff_mine_2014}. \par
Tailings dams are usually constructed by readily available local materials and they are often built of, and/or on tailings material \citep{saad_hydromechanical_2011}. Tailings grain size is highly variable and dependent on the parent rock and the method of extraction. They tend however to be largely gravel-free and clay-free, with sand being more common than silt \citep{kossoff_mine_2014}. Their chemical composition depends on the mineralogy of the ore body, the degree of weathering during storage and the extraction process. Silica, Fe and oxygen display an almost universal presence and seem to be the most abundant elements in tailings \citep{kossoff_mine_2014}. Tailings material is pumped from the mill to the impoundment and is often size differentiated during deposition. The coarser and more porous material settles close to the discharge point, near the embankment, and may be used to extend the structure itself, while the finer fraction (slimes) is carried further away forming an impermeable barrier. This size-differentiated dispersal contributes to the integrity of the dam \citep{kossoff_mine_2014}.
\par
\begin{figure}[H]
		\centering
		\includegraphics[width=\columnwidth, trim={0 0cm  0 0cm}]{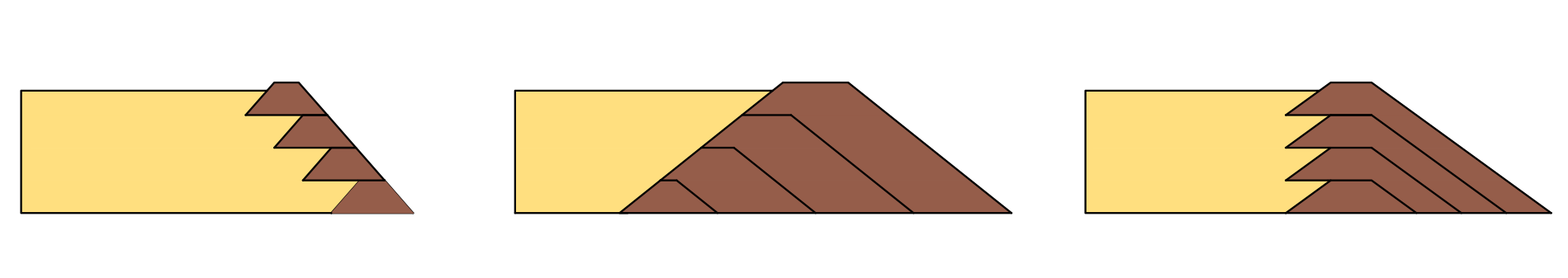}
		\caption{Tailings dam construction method: conceptual illustration. Left to right: Upstream, Downstream, Centreline level raise. Yellow color: Tailings. Brown color: Fill material. }
		\label{up_down_centre}
	\end{figure}

Rather than constructed at once, tailings dams are gradually raised, as the mining activity results to larger capacity demands for the storing reservoir.  Three different methods are used for embankment level raise, namely the upstream, downstream, and centerline method, as illustrated in Figure \ref{up_down_centre}. The brown-colored part in the Figure represents the part of the structure that has to be constructed of coarse and compacted material, and therefore the most expensive part to be built.  The upstream method, requiring the smallest volume of processed fill material, and therefore being the most cost effective, is most widely chosen \citep{lyu_comprehensive_2019}, but it is associated to many major failures. In upstream dams, the additional layers of structure are placed on top of tailings depositions, therefore the construction's safety depends on the integrity of tailings for stability \citep{davies_static_2002}. This type of dam requires greater ongoing scrutiny \citep{martin_considerations_1999}.
	\par

Impoundment failure can be categorized into four basic mechanisms. These main mechanisms are, overtopping, often occurring in inactive structures after a flooding event; localized failure, caused by the presence of a shear band; piping, caused by internal erosion due to seepage; and diffuse failure, triggered by liquefaction of loose tailings material \citep{hamade_geotechnical_2013}.  The common factors in most cases are the importance of the stress and seepage fields on the dam. The seepage field might directly induce instability by erosion, or cause the pore pressure to rise, leading to a reduction of effective stress and shear strength. 
\par
Slope instability -that corresponds to the second failure mechanism, local failure- is one of the most common failure modes \citep{morton_use_2021}. In upstream tailings dams, the risk of instability of the downstream slope should be particularly investigated during design and monitoring. Stability of upstream dams might depend on the stability of the tailings depositions upstream, that support the overlying layers of the embankment. In case the tailings deposited have a high water content, a fast loading rate that allows no time for consolidation may lead to excess pore pressure and loss of resistance to shear failure. The new layers of structure do not provide a stabilizing force to the downstream slope, as is the case with centerline and downstream raising methods. The timescales of loading, i.e. raising the dam level, and of consolidation of the underlying material, are governing the hydromechanical response of the system. \par

Obviously, the stability problem in tailings dams should be treated as a hydro-mechanical problem, as it is the excess of pore pressure that causes the loss of cohesive properties of the material and consequently its local failure. High pore pressures, and high hydraulic gradients in the case of erosive failures (piping), are omnipresent elements in tailings dams failures. Pore pressure monitoring is therefore crucial in the case of tailings dams. \par

	Tailings dams’ safety management involves regular sensing and collection of data describing quantities such as pore pressures and deformations \citep{clarkson_real-time_2020}. The assessment, however, of such large volumes of data is not always straightforward \citep{knutsson_slope_2016},\citep{hui_real-time_2018}.  A common approach is to assess data in terms of trends over time. Expected behaviors are based on previous trends \citep{knutsson_slope_2016}, \citep{vanden_berghe_geotechnical_2011}. This is an appropriate method for monitoring sudden changes that indicate trouble. However, it is not always an appropriate option for non-static structures like tailings dam. During periods when the dam crest level is raised, these structures undergo significant changes. Nonetheless, the induced pore pressures and deformations must remain in the serviceability range. \par

  Monitoring instrumentation and data collection are useful only if they are used in combination with appropriate  numerical models capable of describing the hydro-mechanical coupling that governs the problem \citep{szostak_use_2003}. These models can contribute to identifying undesirable behaviors, thus making sense of the collected data, and establishing appropriate alarm functions. \par 

	A common challenge in developing realistic models is related to highly uncertain parameters, mainly representing material properties \citep{heshmati_r_prediction_2020}. The hydraulic and mechanical properties of the structural and stored materials  are often unknown and may vary over time \citep{villavicencio_estimation_2011}. To treat such problems efficiently, data assimilation may prove useful. In this context, field measurements are used for back-analyses in order to identify realistic values for the properties used in the model. Data assimilation, implies the solution of an inverse problem, that is, a problem that examines multiple solutions for different values of the parameters, and identifies the most realistic parametric values by driving the model output as close to the reality. \par
	
		The use of such methods in a real-time framework presupposes the existence of a numerical model, able to provide responses to many-query problems, faster than the evolution of the physical phenomenon that is analyzed. Reduced Order Models (ROM) prove to be a suitable solution for guaranteeing accurate and fast queries. Model Order Reduction is a set of techniques that aim to enhance a model's computational efficiency by decreasing its dimensionality, that is, the number of degrees of freedom that correspond to the discretized problem. \par
	Data assimilation and inverse problem solving motivate this work, but are not treated in the present paper. This work focuses on implementing Model Order Reduction on the fully coupled hydro-mechanical equations that govern water flow through unsaturated soil. \par
	
	There have been several model reduction methodologies applied to the simulation of porous media flow, such as data-driven models \citep{bao_data-driven_2019} \citep{ghommem_complexity_2016}, Proper Orthogonal Decomposition (POD) based methods \citep{ghommem_complexity_2016} \citep{van_doren_reduced-order_2006} \citep{larion_2020} \citep{badia_coupling_2009}, as well as POD paired with DEIM (Discrete Empirical Interpolation Method) where the non-linear terms are approximated by some form of interpolation, ensuring a large reduction of computational cost \citep{gildin_nonlinear_2013} \citep{esmaeili_generalized_2020}. The work in all the papers mentioned above is motivated by reservoir and petroleum engineering and often refers to multiphase fluid flow. In the aforementioned papers, POD-based reduction is applied to the hydraulic problem alone, and does not concern the hydro-mechanically coupled problem \citep{ghommem_complexity_2016} \citep{gildin_nonlinear_2013} \citep{van_doren_reduced-order_2006} \citep{vermeulen_reduced_2004}.   \par
	The original contribution of this work, lies in the implementation of POD-based reduction for the coupled non-linear problem, considering partial saturation of the porous medium. Coupling of the two equations that govern the mechanical and hydraulic part of the problem, yields a non-linear transient system of equations, that is discretized using the Finite Element Method. Non-linearities are introduced in order to describe partially saturated states for the soil. Partial saturation must be considered in the study of earthdams, as it is a common occurrence in these structures that the water table is located below the dam crest, and therefore only part of the material is saturated.  Solving the discretized system of equations, one can obtain the full-order or high fidelity approximation of the solution to the PDEs. The Reduced Basis method \citep{florentin_adaptive_2012} \citep{maday_reduced-basis_2002} \citep{maday_reduced_2004} \citep{rozza_reduced_2007} represents an instance of model order reduction techniques in which the parametric dependence of the PDE solution is explored by solving the high-fidelity problem a number of times. The resulting set of solutions is explored in a POD framework, in order to find a set of elements, hopefully fewer in number than the dimension of the full-order problem, the linear combination of which can provide a satisfactory approximation of the high-fidelity solution. \par
	To adapt this technique problems in partially saturated soils, the paper is structured as follows. Sections \ref{Equations} to \ref{Methodology} contain a detailed description of the methodologies used for developing the forward FE model and the low-order approximation model using RB. In Section \ref{fenics section} some details on the implementation tool are given. In Section \ref{Application} the accuracy and computational efficiency achieved with ROM are demonstrated solving a problem related to the construction of tailings dams. The results are discussed and an outline of future work that could add value to the present work is given in section \ref{Conclusions}.

	\section{Constitutive relations and governing equations} \label{Equations}
	In this Section, the equations that govern the hydro-mechanically coupled problem of groundwater flow through an unsaturated soil are presented. The equations written here describe various geomechanical problems that feature water flow through unsaturated porous media, and the methodologies developed in this work can be used to solve these problems. 
\subsection{Conservation of linear momentum}

The equation of mechanical equilibrium reads
\begin{equation}
\boldsymbol{\nabla}^\intercal \boldsymbol{\sigma} + \mathbf{b} =\mathbf{0}, \label{Mechanical Equilibrium}
\end{equation} 
where $\boldsymbol{\sigma}$ is the total stress tensor, and $\mathbf{b}$ is the vector of body forces
\begin{equation}
\rm{ \mathbf{b} = \rho(p)\mathbf{g}},
\end{equation}
where $\mathbf{g} = [0,0,-g]^\intercal $ is the gravity acceleration vector in a 3D setting, and $\rho$ is the density of the multiphase medium, comprised of soil particles and water, evaluated as a function of pore water pressure $\rm{p}$, and related to the density of soil particles and water ($\rm{\rho_s, \rho_w}$) according to the relation
\begin{equation}
\rm{ \rho(p) = (1-\eta)\rho_s+ \eta S_e(p) \rho_w= (1-\eta)\rho_s+\Theta(p) \rho_w } 
\end{equation}
where $\eta$ denotes the soil porosity. The volume water content (VWC) $\rm{\Theta(p)}$ and effective degree of saturation $\rm{S_e(p)}$ are evaluated according to a hydraulic model detailed in Section \ref{Water Retention Curve}

In this work the air pressure is considered equal to the atmospheric pressure, as commonly assumed in geotechnics. The constitutive stress is defined as
\begin{equation}
\boldsymbol{\sigma}'= \boldsymbol{\sigma} - \rm{S_e(p)} \mathrm{p}\mathbf{I}. \label{effective stress generic}
\end{equation}
where $\boldsymbol{\sigma'}$ is the tensor of effective stresses, $\mathbf{I}$ the identity matrix and $\rm{S_e}$ the effective degree of saturation \citep{nuth_effective_2008}, or dimensionless water content, which is one of the factors among many \citep{nuth_effective_2008} that have been proposed in order to weight the contributions of the two phases, soil and water, to the total stress. The effective degree of saturation is suction dependent and will be estimated as explained in Section \ref{Water Retention Curve}.\par
Introducing the concept of effective stress $\boldsymbol{\sigma}'$ as defined in equation \eqref{effective stress generic} in the mechanical equilibrium \eqref{Mechanical Equilibrium}, yields,
\begin{equation}
\rm{\boldsymbol{\nabla}^\intercal \left(\boldsymbol{\sigma}'+S_e(p)\rm{p}\mathbf{I}\right) + \mathbf{b} =\mathbf{0} }
\end{equation}
Linear elasticity is assumed for the soil skeleton's response. In that framework, the constitutive stress-strain relation reads
\begin{equation}
\boldsymbol{\sigma}'=\lambda tr \boldsymbol{\varepsilon} (\mathbf{u})\mathbf{I}+2\mu \boldsymbol{\varepsilon} (\mathbf{u})  \label{Hooke's law}
\end{equation}
where $\mathbf{u}$ is the displacement vector, $\boldsymbol{\varepsilon}$ denotes the infinitesimal strain tensor, calculated as $\boldsymbol{\varepsilon}(\mathbf{u})=\frac{\boldsymbol{\nabla}\mathbf{u}+\boldsymbol{\nabla}^\intercal\mathbf{u}}{2}$ , and $\lambda$ and $\mu$ are the Lam\'e elastic moduli. The usual parameters defining the elastic mechanical material characteristics are the Young's modulus $\rm{E}$ and Poisson's ratio $\rm{\nu}$, but the constitutive models are often written in terms of Lamé coefficients. For isotropic 3-dimensional materials and for plane strain conditions in 2 dimensions, the relations among these parameters are,
\begin{align}
\rm{\mu}=\rm{\frac{E}{2(1+\nu)}}, \quad
\rm{\lambda}=\rm{\frac{E\nu}{(1+\nu)(1-2\nu)}}.
\end{align}

Introducing the stress-strain relation, the mechanical equilibrium can be written in the form:
\begin{equation}
\rm{\boldsymbol{\nabla}^{\intercal} \left(\lambda tr \boldsymbol{\varepsilon}(\mathbf{u})\mathbf{I}+2\mu \boldsymbol{\varepsilon}(\mathbf{u}) +S_e(p)\rm{p}\mathbf{I}\right) +\rho(p)\mathbf{g}=\mathbf{0}}  \label{Mechanical equilibrium extended}
\end{equation}
The governing equation \eqref{Mechanical Equilibrium} is accompanied by boundary conditions to formulate a boundary value problem. For the mechanical part of the problem, Dirichlet conditions are used to impose constraints in displacement and Neumann conditions to apply a load.\par
Let  $\rm{ \Gamma_D^u}$, $\rm{ \Gamma_N^u}$ and be two partitions of the boundary $\rm{\partial \Omega}$ of the domain $\rm{\Omega}$ on which Dirichlet and Neumann boundary conditions are applied respectively. The boundary conditions are:
\begin{align}
\rm{\boldsymbol{\mathbf{u}}} & = \hat{\mathbf{u}} \quad \text{on} \ \rm{\Gamma_D^u}  \label{BC 1} \\
\rm{\boldsymbol{\sigma}\cdot \mathbf{n} } &=\rm{ \hat{ \mathbf{t}} } \quad \text{on} \ \Gamma_N^u   \label{BC 2}
\end{align}
where $\rm{\mathbf{n}}$ is the outward pointing unit normal vector and $\rm{\hat{\mathbf{t}}}$ is the surface traction force.

\subsection{Water mass conservation}
Considering the mass balance of pore fluids leads to the continuity equation for flow, stating that the water outflow from a representative elementary volume is equal to the changes in mass concentration. Neglecting the deformations of solid particles due to effective stress and pore pressure, and the density gradients of water, introducing the Darcian definition for fluid velocity stated in equation \eqref{Darcy's law}, the strong form of the continuity equation reads
\begin{equation}
\rm{   \boldsymbol{\nabla}^\intercal\big[  \frac{k(p)} {\gamma_w} ( \boldsymbol{\nabla}\mathrm{p} + \mathbf{b}_w)  \big]  + \bigg( C(p) - \frac{ \Theta(p) }{K_w} \bigg)  \dot{\mathrm{p}}   =  \Theta(p)  \boldsymbol{\nabla}\cdot\dot{ \mathbf{u}}   }, \label{Fluid Strong form final}
\end{equation}
where $\rm{\mathbf{b}_w = \rho_w\mathbf{g}}$ are the body water forces and $\rm{K_w}$ is the water bulk modulus. The hydraulic conductivity  $ \rm{k(p)}$, specific moisture capacity $\rm{C(p)}$, and volumetric water content (VWC) $\rm{\Theta(p)}$ are estimated using the soil water retention relations that are presented  in Section \ref{Water Retention Curve}.\par
The governing equation describing water flow is transient, with two time dependent terms. One contains the time derivative of pore pressure $\dot{\mathrm{p}}$, and the other, a coupling term which accounts for the change in porosity due to the overall compression of the soil structure, and contains the time derivative of the volumetric strain $\dot{ \mathbf{u}}$ of the medium.
The specific discharge, or fluid velocity $\mathbf{q}$ can be related to the pore water pressure gradient $\rm{\boldsymbol{\nabla}p}$ according to Darcy's law, which for an isotropic material takes the form
\begin{equation}
\rm{ \mathbf{q}=-\frac{k(p)}{\gamma_w} ( \boldsymbol{\nabla}\mathrm{p} +  \mathbf{b}_w }) \label{Darcy's law},
\end{equation}\par
where $\rm{k}$ is the hydraulic conductivity of the multiphase medium measured in (m/s), $\rm{\gamma_w}$ is the specific weight of water and $\rm{\mathbf{b}_w}$ is the vector of fluid body forces.

Typical boundary conditions that arise in the case of earthfill dams may be either of Dirichlet, Newmann or Robin type. For the flow equation, Dirichlet conditions may be used to prescribe a known hydraulic head, Neumann conditions for a known outflow, inflow or a hydraulically closed (impervious) boundary.\par
A particular case arises in the description of a seepage face, which occurs when a water table touches an open downstream boundary \citep{pinyol_rapid_2008} \citep{gerard_study_2009}. The length of the seepage surface is pressure-dependent \citep{alonso_review_2005}, and can be prescribed as a non-linear Robin condition.\par 
Let  $\rm{  \Gamma_D^p}$, $\rm{\Gamma_N^p}$ and $\rm{ \Gamma_R^p}$ be three partitions of the boundary $\rm{\partial \Omega}$ of the domain $\rm{\Omega}$ on which Dirichlet, Neumann and Robin boundary conditions are applied respectively. The boundary conditions are:
\begin{align}
\rm{p} & = \hat{\rm{p}} \quad \text{on} \ \rm{\Gamma_D^p} \label{BC 3} \\
\rm{ \mathbf{q} \cdot \mathbf{n}}& =\rm{ \hat{q} } \quad \text{on} \ \Gamma_N^p \label{BC 4} \\
\rm{ \mathbf{q} \cdot \mathbf{n} }& =\rm{\langle \beta p \rangle} \quad \text{on} \ \Gamma_R^p \label{BC 5}
\end{align}
where $\rm{\mathbf{n}}$ is the outward pointing normal vector and $\rm{\hat{q}}$ is the fluid flux on the boundary. Equation \eqref{BC 5} refers to the seepage condition, where the non-linear function of $\rm{p}$ that is denoted with angular brackets prescribes a flux that is equal to $\rm{\beta p}$, when $\rm{p} >0 $ and vanishes for negative pressure. The coefficient $\beta $ depends on the hydraulic conductivity and geometry of the domain and defines the water runoff on a boundary in seepage conditions. This is a non-linear Robin type condition.

\subsection{Soil water characteristics}\label{Water Retention Curve}
The most commonly used hydraulic model for the water content - pore water pressure relation in unsaturated soils is the one proposed by Van Genuchten \citep{van_genuchten_closed-form_1980}. The effective saturation $\rm{S_e}$ -or dimensionless water content- is given by 
\begin{equation}
\rm{S_e(p)}= \left\{ 
\begin{array}{ll}
\rm{\frac{1}{[1+(\alpha|\frac{p}{\gamma_w}|)^{\frac{1}{1-m}}]^m}} & \rm{p < 0}\\
\rm{1} & \rm{p \ge 0}\\
\end{array}
\right.\label{V.Genucht}
\end{equation}
where $\rm{\alpha}$ is a parameter related to the air entry value of the soil and $\rm{m}$ is a curve fitting parameter. The upper branch of this equation describes a sigmoid curve which is called a water-retention curve. The VWC is then given by 
\begin{equation} \rm{\Theta(p)=S_e(p)(\Theta_s-\Theta_r)+\Theta_r} \label{VWC} \end{equation}
where $\rm{\Theta_s , \Theta_r}$ are soil characteristics: the VWC for fully saturated conditions, and the residual VWC. 
Differentiation of equation \eqref{VWC} with respect to pore water pressure gives
\begin{equation}\rm{C(p)= \frac{\partial \Theta(p)}{ \partial p} = \frac{-\alpha m (\Theta_s-\Theta_r)}{1-m} S_e(p)^{1/m}(1-S_e(p)^{1/m})^m}.\end{equation}
The relation between the hydraulic conductivity of the soil-water system and the pore water pressure as proposed by van Genuchten \citep{van_genuchten_closed-form_1980} reads,
\begin{equation}\rm{k(p)=k_{s}\sqrt{S_e(p)}\big[1-\big(1-S_e(p)^{1/m}\big)^m\big]^2} \end{equation}
where $\rm{k_s}$ is the hydraulic conductivity for saturated conditions.

	\section{Model reduction methodology} \label{Methodology}
\subsection{Finite Element Method for hydro-mechanical groundwater flow problems in unsaturated conditions}
Multiplying equation \eqref{Mechanical equilibrium extended} with a vector test function $\mathbf{v}$, integrating over the domain $\Omega$,  and applying the Green-Gauss theorem, the variational form is obtained:
\begin{equation} \int_{\Omega} \boldsymbol{\sigma}':\boldsymbol{\varepsilon}{(\mathbf{v})}\rm{dx} - \int_{\Omega} S_e \rm{p}\mathbf{I}:\boldsymbol{\varepsilon}{(\mathbf{v})}\rm{dx}  = \int_{\Omega} \mathbf{b}\mathbf{v}\rm{dx}+	\int_{\Gamma_{N}^\mathbf{u}}(\boldsymbol{\sigma}\cdot\mathbf{n})\mathbf{v}\rm{ds} \label{Mech Variational} \end{equation}
The equation is discretized applying the Galerkin approach. The discretized equation reads
\begin{equation}
\mathbf{K}\mathbf{U} -  \mathbf{Q}\mathbf{P} =  \mathbf{f_u} \label{Discretized Mech}
\end{equation}
where,
\begin{align*}
\mathbf{K} &=  \rm{\int_{\Omega}  \boldsymbol{\nabla}\mathbf{N_u}^\intercal \mathbf{D}_{\text{el}}   \boldsymbol{\nabla}\mathbf{N_u} dx  }\\
\mathbf{Q} &= \rm{  \int_{\Omega}  \boldsymbol{\nabla}\mathbf{N_u}^\intercal S_e \mathbf{N}_p dx}\\
\mathbf{f_u} & = \rm{  \int_{\Omega} \mathbf{N_u} \mathbf{b} dx + \int_{\Gamma_{N}^u} \mathbf{N_u} \hat{\mathbf{t}} ds }. \end{align*}
$\mathbf{N_u}, \mathbf{N}_p$ are displacement and water pressure shape function matrices respectively, and $\mathbf{U}$ and $\mathbf{P}$ are unknown nodal value vectors for displacement and pressure, such that $\mathbf{u} \approx \mathbf{N_u} \mathbf{U}$ and $ \rm{p} \approx  \mathbf{N}_p \mathbf{P} $ and $\mathbf{D}_\text{el}$ is the elastic stress-strain matrix.

Similarly, multiplying equation \eqref{Fluid Strong form final} by scalar test function $\rm{w}$, the variational form of the water flow equation is obtained, assuming no inflow or outflow in the domain:
\begin{equation} \rm{ \int_{\Omega}\frac{k} {\gamma_w}  \boldsymbol{\nabla}\mathrm{p} \cdot \boldsymbol{\nabla} w dx +  \int_{\Gamma_{R}^p} \langle \beta  p \rangle  w ds + \int_{\Omega}\Theta \boldsymbol{\nabla}\cdot\dot{ \mathbf{u}}wdx - \int_{\Omega}\big(C- \frac{\Theta }{K_w} \big)\dot{p}wdx  = }
\rm{   \int_{\Omega}\frac{k} {\gamma_w} \mathbf{b_w} \cdot \boldsymbol{\nabla} wdx  } \label{Hyd Variational} \end{equation} 

\begin{equation} \rm{ \mathbf{H}\mathbf{P}+ \mathbf{C}\dot{\mathbf{U}} - \mathbf{S}\dot{\mathbf{P}} =  \mathbf{f}_p}  \label{Discretized Hyd} \end{equation}
	\begin{align*}
	\rm{\mathbf{H}}&=\rm{\int_{\Omega}\boldsymbol{\nabla}\mathbf{N}_p^\intercal\frac{k}{\gamma_w}\boldsymbol{\nabla}\mathbf{N}_pdx+  \int_{\Gamma_{R,p>0}^p} \mathbf{N}_p^\intercal \beta \mathbf{N}_p ds}\\
	\rm{  \mathbf{C}} & \rm{= \int_{\Omega}\mathbf{N}_p^\intercal \Theta \boldsymbol{\nabla} \mathbf{N_u}dx}   \\
 \rm{\mathbf{S} }&\rm{= \int_{\Omega}\mathbf{N}_p^\intercal\big(C- \frac{\Theta }{K_w} \big) \mathbf{N}_pdx} \\
 \rm{  \mathbf{f}_p} &\rm{=\int_{\Omega}\boldsymbol{\nabla}\mathbf{N}_p \frac{k}{\gamma_w}\mathbf{b}_w dx }    \end{align*}

To solve equations \eqref{Discretized Hyd} and \eqref{Discretized Mech}, time stepping is implemented by a generalized $\theta$-scheme, which approximates $\rm{\mathbf{X}^\intercal = [\mathbf{U} \ \mathbf{P}]^\intercal}$ at time $\rm{i+\theta}$ as
	\begin{equation}
\rm{\dot{\mathbf{X}}^{i+\theta} \simeq  \frac{\mathbf{X}^{i+1}-\mathbf{X}^{i}}{\Delta t}}
, \quad \mathbf{X}^{i+\theta}\simeq (1-\theta)\mathbf{X}^{i}+\theta \mathbf{X}^{i+1}, \label{theta scheme}
\end{equation}
where $\rm{\Delta t}$ is the time step and $\rm{i+1}$ denotes the current time step. Parameter $\rm{\theta}$ takes values in $\rm{[0,1]}$. \par
Operators $\mathbf{Q}$, $\mathbf{H}$, $\mathbf{C}$, $\mathbf{S}$, all depend on the pressure state, therefore they must be re-evaluated at each time instance. The same applies for force vectors $\mathbf{f_u}$ and $\rm{\mathbf{f}_p}$.  Solving equations \eqref{Discretized Hyd} and \eqref{Discretized Mech} at time $\rm{i+\theta}$, the system reads:
\begin{align}
\rm{\mathbf{K}\mathbf{U}^{i+\theta} -  \mathbf{Q}^{i+\theta}\mathbf{P}^{i+\theta} } &= \rm{ \mathbf{f_u}^{i+\theta} },\label{Disc Mech_i+theta} \\
\rm{ \mathbf{H}^{i+\theta}\mathbf{P}^{i+\theta}+ \mathbf{C}^{i+\theta}\frac{\mathbf{U}^{i+1}-\mathbf{U}^{i}}{\Delta t} - \mathbf{S}^{i+\theta}\frac{\mathbf{P}^{i+1}-\mathbf{P}^{i}}{\Delta t}} &=  \rm{\mathbf{f}_p^{i+\theta}}.\label{Disc Hyd_i+theta}
\end{align}
The time stepping scheme as presented in equation \eqref{theta scheme} is used for the approximation of operators and vectors that depend on the pressure state. Hence, the operator $\mathbf{Q}$ at time $\rm{i+\theta}$ is approximated as,
\begin{equation}
\rm{\mathbf{Q}^{i+\theta}\simeq (1-\theta)\mathbf{Q}^{i}+\theta \mathbf{Q}^{i+1}}.
\end{equation}
Other operators, $\mathbf{Q}^{i+\theta}$, $\mathbf{H}^{i+\theta}$, $\mathbf{C}^{i+\theta}$, $\mathbf{S}^{i+\theta}$, $\mathbf{f_u}^{i+\theta}$ and $\rm{\mathbf{f}_p}^{i+\theta}$ are approximated similarly. Introducing these approximations to equations \eqref{Disc Mech_i+theta} and \eqref{Disc Hyd_i+theta}, the  fully coupled discretized system using a monolithic approach reads, \par
\begin{equation}
\rm{\begin{bmatrix}\hat{\mathbf{K}}& \hat{\mathbf{Q}} \\ \hat{\mathbf{C}} &\hat{\mathbf{H}}\end{bmatrix} \begin{bmatrix}\mathbf{U}\\ \mathbf{P}  \end{bmatrix}^{i+1}  = \begin{bmatrix} \hat{\mathbf{f_u}} \\ \hat{\mathbf{f}_{\rm{p}}} \end{bmatrix} }, \label{Coupled System}
\end{equation}
where the components $\hat{\mathbf{K}},\hat{\mathbf{Q}},\hat{\mathbf{C}},\hat{\mathbf{H}} $ of the global stiffness matrix and the force vector are evaluated as,

\begin{align}
\rm{ \hat{\mathbf{K}}} &=\rm{ \theta   \mathbf{K}},\\
\rm{ \hat{\mathbf{Q}} }&= \rm{ -\theta (1-\theta)\mathbf{Q}^i - \theta^2   \mathbf{Q}^{i +1}     }, \\
\rm{ \hat{ \mathbf{f_u} } }& =\rm{ - \mathbf{K}(1-\theta) \mathbf{P}^i + [\theta(1-\theta)\mathbf{Q}^{i+1}+  (1-\theta)^2\mathbf{Q}^i ] \mathbf{U}^i +(1-\theta)\mathbf{f_u}^i +\theta\mathbf{f_u}^{i+1} },\\ 
\rm{ \hat{\mathbf{C}} }&= \rm{\theta \mathbf{C}^{i+1} +(1-\theta)\mathbf{C}^i },\\
\rm{ \hat{\mathbf{H}} } &= \rm{ \Delta t (1-\theta)\theta \mathbf{H}^{i} + \Delta t\theta^2 \mathbf{H}^{i+1} -(1-\theta) \mathbf{S}^i - \theta \mathbf{S}^{i+1}}, \\
\rm{ \hat{\mathbf{f}}_p} &=  \rm{-[ \Delta t (1-\theta)^2 \mathbf{H}^{i} + \Delta t(1-\theta)\theta \mathbf{H}^{i+1} \theta +(1-\theta) \mathbf{S}^i + \theta \mathbf{S}^{i+1} ]\mathbf{P}^i +[ \theta\mathbf{C}^{i+1} +(1-\theta)\mathbf{C}^i ] \mathbf{U}^i + \Delta t (1-\theta) \mathbf{f}_p^i + \theta\mathbf{f}_p^{i+1} }.
\end{align}
In this work, the non-linear system of equations is solved using a Picard iterative scheme.

\subsection{Model Order Reduction: The Reduced Basis Method }

In data assimilation problems, where the value of some parameters needs to be determined based on the available information, many queries have to be done to the numerical model. If it is based on a full-order approach (e.g. FE) the computational cost might become prohibitive.\par
The Reduced Basis (RB) \citep{quarteroni_reduced_2016} \citep{hesthaven_certified_2016}\citep{florentin_adaptive_2012} method tries to create a small basis that is able to represent the family of the solutions spanned by the parameters' variations. The simplest method to create the RB is to solve the full-order FE problem at a set of parametric values that capture the overall behavior of the solution. Each one of these samples is usually called a snapshot.\par
Once the parametric space has been sampled, the RB is constructed by orthonormalizing the set of snapshots and discarding those with amplitudes smaller than a certain threshold. This process is usually called off-line, as it is done once, i.e. it can be seen as a pre-process of the data-assimilation.\par
Once the RB is ready, the solution of the problem for any point in the parametric space can be obtained as a linear combination of the members of the RB. Therefore, the computational cost is largely reduced as the number of unknowns to determine is usually several orders of magnitude smaller than the size of the original FE problem. The small  problem size allows for a very fast solution that can be done repetitively within the assimilation of data. This fast solution is usually called the on-line phase of the RB method.\par

In the following, the parameter vector is denoted by $\boldsymbol{\mu} \in \mathcal{P} \subset \mathbb{R}^P$ where the parameter space  $\mathcal{P} $  represents a closed and bounded subset of the Euclidean space $\mathbb{R}^P, P<1$. The field variable given by the Finite Element solution of a parametrized PDE can be seen as a map $\mathbf{x} \ : \ \mathcal{P} \rightarrow V$, that to any $ \boldsymbol{\mu} \in \mathcal{P} $ associates the solution $\mathbf{x}(\boldsymbol{\mu})$ belonging to a suitable functional space V.  \par
The full-order approximation of a PDE for a given $\boldsymbol{\mu} \in \mathcal{P} $ can be represented in the generic form
\begin{equation}
\mathbf{A}(\boldsymbol{\mu})\mathbf{x}(\boldsymbol{\mu})=\mathbf{f}(\boldsymbol{\mu})
\end{equation}
where $\mathbf{A}(\boldsymbol{\mu}) \in \mathbb{R}^{N_h \times N_h}$ and $\mathbf{f}(\boldsymbol{\mu}) \in \mathbb{R}^{N_h}$ are a  $\boldsymbol{\mu}$ -dependent matrix and vector respectively, representing the stiffness matrix and the force vector. The system has $N_h$ degrees of freedom. \par
The key idea of RB, is to replace this system with another one, of lower dimension $N_r<N_h$ \citep{quarteroni_reduced_2016} . For any given $\boldsymbol{\mu} \in \mathcal{P} $, the solution field is approximated as $\mathbf{x}(\boldsymbol{\mu}) \approx \mathbf{B}\boldsymbol{\alpha}(\boldsymbol{\mu})$ and the low-order system reads:
\begin{equation}
 \mathbf{B}^{\intercal}\mathbf{A}(\boldsymbol{\mu})\mathbf{B}\boldsymbol{\alpha}(\boldsymbol{\mu})=\mathbf{B}^{\intercal}\mathbf{f}(\boldsymbol{\mu}) \label{projection}
\end{equation}
where $( \mathbf{B}^{\intercal}\mathbf{A}(\boldsymbol{\mu})\mathbf{B}) \in \mathbb{R}^{N_r \times N_r}$, $(\mathbf{B}^{\intercal}\mathbf{f}(\boldsymbol{\mu}) ) \in \mathbb{R}^{N_r}$ and $\boldsymbol{\alpha}(\boldsymbol{\mu})$ is the reduced vector of degrees of freedom. The form $\mathbf{B}\boldsymbol{\alpha}(\boldsymbol{\mu})$  represents the approximation of the high-fidelity solution $\mathbf{x}(\boldsymbol{\mu})$, in the low-order space $\mathbb{R}^{N_r}$, where $\mathbf{B} \in \mathbb{R}^{N_h \times N_r}$ is an $\boldsymbol{\mu}$-independent transformation matrix, the columns of which collect the reduced basis vectors.  
\par 

For time-dependent problems, like the one at hand, the full-order PDE approximation is written in a general form as

\begin{equation}
\mathbf{M}(t; \boldsymbol{\mu})\dot{ \mathbf{x}}(t; \boldsymbol{\mu}) + \mathbf{A}(t;\boldsymbol{\mu})\mathbf{x}(t;\boldsymbol{\mu})=\mathbf{f}(t;\boldsymbol{\mu}),
\end{equation}
where $\mathbf{A}(t; \boldsymbol{\mu}), \mathbf{M}(t; \boldsymbol{\mu}) \in \mathbb{R}^{N_h \times N_h}$ are time and parameter-dependent matrices and  $\mathbf{f}(t;\boldsymbol{\mu}) \in \mathbb{R}^{N_h}$ is a vector of  $\boldsymbol{\mu}$ and time-dependent data. Considering the approximation of the time derivative $\dot{ \mathbf{x}}(t; \boldsymbol{\mu}) \simeq  \frac{\mathbf{x}^{i+1}-\mathbf{x}^{i}}{\Delta t} $, the reduced-order approximation of the PDE for any time level $t^i = i\Delta t$, ($\Delta t>0$ being the time step) reads \citep{hesthaven_certified_2016},
\begin{equation}
 \mathbf{B}^{\intercal}\left( \frac{1}{\Delta t} \mathbf{M}(t; \boldsymbol{\mu}) + \mathbf{A}(t;\boldsymbol{\mu})\right)  \mathbf{B}\boldsymbol{\alpha}^{i}(\boldsymbol{\mu}) =  \mathbf{B}^{\intercal} \left(\frac{1}{\Delta t} \mathbf{M}(t; \boldsymbol{\mu}) \mathbf{B}\boldsymbol{\alpha}^{i-1}(\boldsymbol{\mu}) + \mathbf{f}(t;\boldsymbol{\mu})\right).
\end{equation}

The basis creation in the presence of non-homogeneous Dirichlet boundary conditions is treated here by isolating the known boundary degrees of freedom from the unknown values to determine \citep{HOANG201896}. Thus the reduced approximation of the solution $\mathbf{x}(\boldsymbol{\mu})$ reads,
\begin{equation}
\mathbf{x}(\boldsymbol{\mu})\approx \begin{bmatrix} \mathbf{B} \\ \mathbf{0} \end{bmatrix}\boldsymbol{\alpha}(\boldsymbol{\mu}) + \begin{bmatrix} \mathbf{0} \\ \hat{\mathbf{x}}  \end{bmatrix} 
\end{equation}
and the reduced problem becomes homogeneous. This guarantees the exact fulfillment of the Dirichlet boundary conditions.
The reduced problem now reads,
\begin{equation}
\begin{bmatrix} \mathbf{B} \\ \mathbf{0} \end{bmatrix}^{\intercal}\mathbf{A}(\boldsymbol{\mu}) \begin{bmatrix} \mathbf{B} \\ \mathbf{0} \end{bmatrix}\boldsymbol{\alpha}(\boldsymbol{\mu}) = \begin{bmatrix} \mathbf{B} \\ \mathbf{0} \end{bmatrix}^{\intercal} \mathbf{f}(\boldsymbol{\mu}) - \begin{bmatrix} \mathbf{B} \\ \mathbf{0} \end{bmatrix}^{\intercal}\mathbf{A}(\boldsymbol{\mu}) \begin{bmatrix} \mathbf{0} \\ \hat{\mathbf{x}} \end{bmatrix}. \label{Treating BCS}
\end{equation}
Thus the final solution necessarily respects the Dirichlet boundary conditions.\par 
For the problem at hand, two separate low-order bases are built to approximate each unknown field \citep{quarteroni_reduced_2014} \citep{ortegagelabert_fast_2020}. Transformation matrices $\rm{\mathbf{B_u}, \mathbf{B}_p}$ correspond to the displacement and pressure fields respectively.\par
In the following, the indicator of dependence of operators on the parameter vector $(\boldsymbol{\mu})$ has been omitted for clarity. The unknown vectors are approximated as,
\begin{align}
\mathbf{U}&\approx \mathbf{B_u} \boldsymbol{\alpha}_{\mathbf{u}}, \\
\mathbf{P}&\approx \mathbf{B}_{\rm{p}} \boldsymbol{\alpha}_{\mathrm{p}},
\end{align}
and the reduced dimensional system to be solved, at time step $\rm{i+1}$ reads,
\begin{equation}\label{RB System}
\begin{bmatrix}  \mathbf{B_u}^\intercal \hat{\mathbf{K}}\mathbf{B_u} & \mathbf{B_u}^\intercal \hat{ \mathbf{Q}} \mathbf{B_p}\\
\mathbf{B_p}^\intercal \hat{ \mathbf{C}}\mathbf{B_u} & \mathbf{B_p}^\intercal\hat{ \mathbf{H}}\mathbf{B_p}\end{bmatrix}
\begin{bmatrix} \boldsymbol{\alpha}_\mathbf{u}\\ \boldsymbol{\alpha}_\mathrm{p} \end{bmatrix}^{\rm{i+1}}
=
\begin{bmatrix} \mathbf{B_u}^\intercal\hat{\mathbf{f}}_{\mathbf{u}}  \\ \mathbf{B_p}^\intercal \hat{\mathbf{f}}_{\rm{p}}  \end{bmatrix}.
\end{equation}
The unknowns in this new system are vectors $ \boldsymbol{\alpha}_\mathbf{u}$ and  $\boldsymbol{\alpha}_\mathrm{p}$ that contain the coefficients for linearly combining the elements in the reduced bases, to approximate the high-fidelity solution for any parametric value. \par

\subsubsection*{Constructing the Reduced Basis} \label{Constructing RB}
We denote $\mathcal{M}$ the solution map, or solution manifold of the high-fidelity problem. $\mathcal{M}$ represents the set of solutions $\mathbf{x}(\boldsymbol{\mu})$ for all parameters $\boldsymbol{\mu} \in \mathcal{P} \subset \mathbb{R}^P$, defining a map:
\begin{equation}
\mathcal{M} = \{ \mathbf{x}(\boldsymbol{\mu}) \in V \ \colon \ \boldsymbol{\mu} \in \mathcal{P} \subset \mathbb{R}^P \} \label{Manifold}
\end{equation}
The idea behind RB is to sample the solution manifold by taking snapshots, and use these snapshots to create the reduced space in which the reduced solution is sought. To achieve this, we start from a set of $N_s$ high-fidelity solutions
that are stored in a matrix $\mathbf{M} \in \mathbb{R}^{N_h \times N_s}$, as
\begin{equation}
\mathbf{M}=   [\mathbf{x}^1 ...\mathbf{x}^{N_s}].
\end{equation}
That set of solutions, if well selected, contains the information necessary to describe the parametric dependency of the solution with an acceptable accuracy.\par 
The Proper Orthogonal Decomposition (POD) will be used for the estimation of the reduced basis functions. The singular value decomposition of the matrix $\mathbf{M} \in \mathbb{R}^{N_h \times N_s}$ yields the product representation $\mathbf{M}=\mathbb{U}\boldsymbol{\Sigma}\mathbb{V}^{\intercal}$. The columns of left singular matrix $\mathbb{U}^{N_h \times N_h}$ are orthonormalized vectors that contain information on the parametric dependency of the snapshots. The matrix $\boldsymbol{\Sigma}$ is a diagonal matrix that contains the singular values $\sigma_1,\sigma_2...,\sigma_P, \ P=min\{N_h,N_s\}$. Extracting the $N_r$ first columns of $\mathbb{U}$ will yield the transformation matrix $\mathbf{B}^{N_h \times N_r}$. The number $N_r$ of left singular vectors that are kept may be evaluated based on the singular value corresponding to each vector. The singular values provide a measure of the information of the matrix $\mathbf{M}$ that is captured by each vector. Hence one might extract the first $N_r$ columns of matrix $\mathbb{U}$ that correspond to the $N_r$ largest singular values, and carry the most essential information. \par
Two separate snapshot matrices, each containing the degrees of freedom that correspond to each field are created. The problem is transient, so the snapshots consist of solution vectors for each time step, saved serially in the snapshot matrices as seen below.
		\begin{equation} \label{Mu}
\mathbf{M}_{\mathbf{u}}=\begin{bmatrix}\\  \begin{bmatrix} \\ \mathbf{u}_1 & \mathbf{u}_2 &... & \mathbf{u}_{N_t} \\ \\ \end{bmatrix}_1 
\begin{bmatrix}\\  \mathbf{u}_1 & \mathbf{u}_2 &... & \mathbf{u}_{N_t} \\  \\ \end{bmatrix}_2 ...
\begin{bmatrix}\\  \mathbf{u}_1 & \mathbf{u}_2 &... & \mathbf{u}_{N_t} \\ \\ \end{bmatrix}_{N_{s}}\\ \\ \end{bmatrix}
\end{equation}

\begin{equation} \label{Mp}
\mathbf{M}_{\rm{p}}=\begin{bmatrix}\\  \begin{bmatrix} \\ \mathrm{p}_1 & \mathrm{p}_2 &... & \mathrm{p}_{N_t}  \\ \\ \end{bmatrix}_1 
\begin{bmatrix}\\  \mathrm{p}_1 & \mathrm{p}_2 &... & \mathrm{p}_{N_t} \\  \\ \end{bmatrix}_2 ...
\begin{bmatrix}\\  \mathrm{p}_1 & \mathrm{p}_2 &... & \mathrm{p}_{N_t} \\ \\ \end{bmatrix}_{N_{s}}\\ \\ \end{bmatrix}
\end{equation} 
$N_t$ is the number of time steps that constitute the snapshots, and $N_s$ is the number of snapshots taken.\par
Singular value decomposition is applied to each matrix $\mathbf{M_u}=\mathbb{U}_{\mathbf{u}}\boldsymbol{\Sigma}_{\mathbf{u}}\mathbb{V}_{\mathbf{u}}^{\intercal}$ , $\mathbf{M}_{\rm{p}}=\mathbb{U}_{\rm{p}} \boldsymbol{\Sigma}_{\mathrm{p}}\mathbb{V}_{\rm{p}}^{\intercal}$, thus obtaining left singular matrices $\mathbb{U}_{\mathbf{u}}$ and $\mathbb{U}_{\rm{p}} $, to be truncated based on singular values, as mentioned above.

	\section{FeniCS computing platform} \label{fenics section}
	The model was developed in FEniCS platform \citep{aln_fenics_2015}. FEniCs Project is a collection of free and open-source software components with the common goal to enable automated solutions of differential equations. The components provide scientific computing tools for working with computational meshes, finite element variational formulations of ordinary and partial differential equations, and numerical linear algebra.  It enables users to quickly translate physical models into efficient FEM code easily. The high-level interface, which is in Python or C++ is intuitive and simple for basic use, while also providing options for advanced use. \par
	It provides tools such as languages that allow the user to declare variational forms and compilers that automatically translate these forms. It comes with built-in libraries that offer ready-made Finite Element operators. For example, in FEniCS, a stiffness matrix can be assembled by calling a single command, if the corresponding weak form has been declared using the built-in languages. In the example in Listing \ref{Fenics code}, a stiffness matrix is computed that corresponds to a PDE system that models the deformation of an elastic body. The weak form of the governing equations must be supplied in a symbolic form using UFL (Unified Form Language), while the other components of FEniCS can automatically handle the mesh generation, function space definition(Finite element Automatic Tabulator FIAT), finite element assembly (DOLFIN) and system solving that is supported by several linear algebra packages interfaced in FeniCS.  \par 
	\begin{lstlisting}	[caption={Assembly of stiffness matrix for linear elasticity problem with FEniCS},label={Fenics code}, language=Python, numbers=none]
	from fenics import *
	
	u = TrialFunction(V)
	v = TestFunction(V)
	d = u.geometric_dimension() 
	
	def epsilon(u):
	return 0.5*(nabla_grad(u) + nabla_grad(u).T)
	
	def sigma(u):
	return lambda_*nabla_div(u)*Identity(d) + 2*mu*epsilon(u)
	
	a = inner(sigma(u), epsilon(v))*dx
	
	A = assemble(a)

	\end{lstlisting}
	The trial and test functions are then declared, then constitutive equations are declared using UFL operators such as $ \text{nabla\_grad, nabla\_div} $. The weak form for the left hand side of the equation is declared, again using the UFL operator $ \text{inner}$, and finally the stiffness is automatically assembled with a call to the corresponding DOLFIN routine.
	
	There are even more sophisticated routines that facilitate the declaration and resolution of any PDE-restricted problem including transient or non-linear problems that must, however, be adjusted to the specific application, in this case geotechnical.\par

   \section{Reduced Order Model for predictive monitoring of Tailings Dam} \label{Application}

\subsection{Problem Setup}

In this Section, a ROM of a tailings dam is created to simulate a problem that corresponds to level raise. 
A complete study of the structure's integrity in such conditions, would imply modelling the tailings deposit, as well as the embankment, and considering a possible spatial variation in the mechanical and hydraulic properties. This would be necessary given that the failure surface often appears partly in the impoundment. \par 
In this paper, only a first approach to address the problem is undertaken in order to demonstrate the high level of accuracy that has been achieved using Model Order Reduction to solve a complex, hydro-mechanically coupled, transient non-linear problem, and discuss the contribution of such technology to real-time predictive monitoring. In that line, some simplifying modeling assumptions have been adopted. The impoundment is only treated as a load applied to the embankment and the water table upstream is considered stable even after the loading. The dam is considered to be founded on an impervious layer. \par

\begin{figure}[H] 
	\centering
	\begin{subfigure}[H]{0.65\columnwidth}
		\includegraphics[width=1.2\columnwidth, trim={1cm 0cm  0 0cm}]{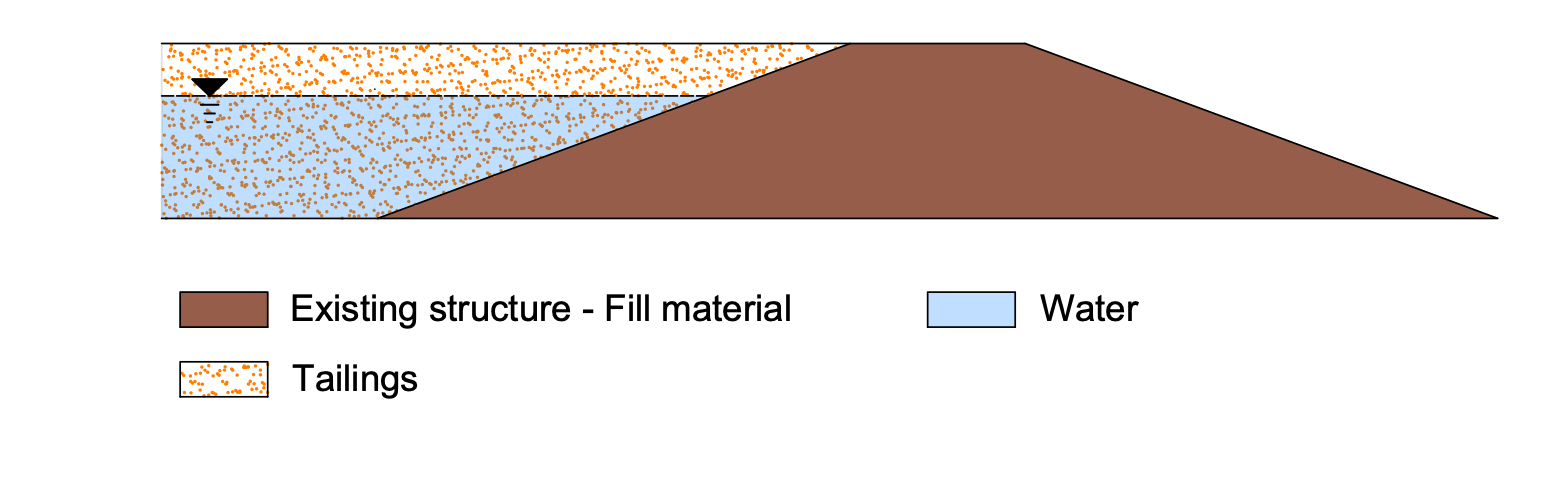}
		\caption{}
		
	\end{subfigure}
	
	\begin{subfigure}[H]{0.65\columnwidth}
		\includegraphics[width=1.2\columnwidth, trim={1cm 0cm  0 0cm}]{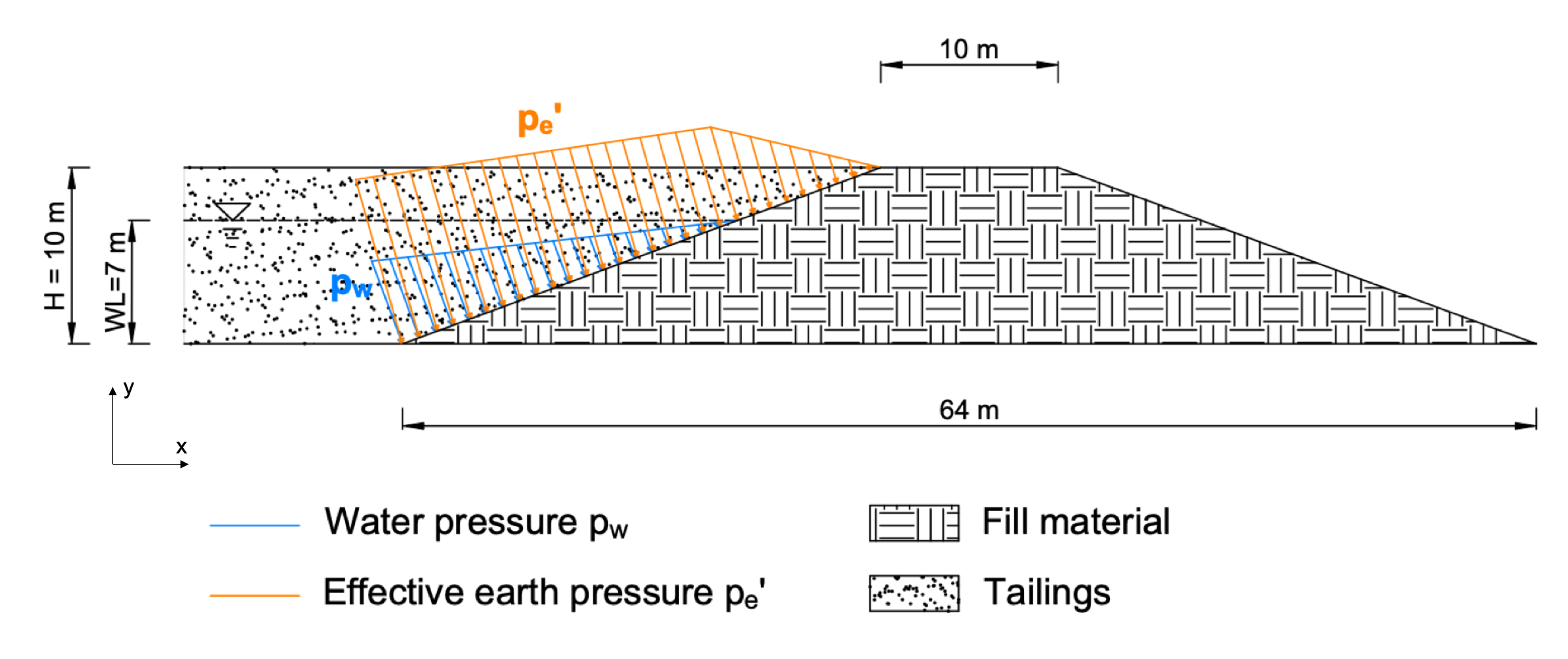}
		\caption{}
		
	\end{subfigure}
	
	\caption{Initial conditions. (a) Conceptual sketch (b) Loads due to water and tailings as modeled }
	\label{Initial Conditions pic}
\end{figure}

 The modeled domain corresponds to the original embankment, while the deposited tailing material upstream, as well the added material on top are modeled as mechanical loads. Initial conditions represent a steady state reached for a known water level imposed upstream, as shown in Fig. \ref{Initial Conditions pic}.

Loads $\rm{p_w}$ and $\rm{p_e}$ occur due to water and tailings deposit respectively. They are evaluated as,
\begin{align}
&\rm{p_w = \gamma_w \times (\rm{WL} - y)}\\
&\rm{ p_{ex}} =
\left\{
\begin{array}{ll}
\rm{K_a\big[\gamma_t(H-WL) + (\gamma_t - \gamma_w)(WL-y)\big] } & \rm{\mbox{if } y \leq WL} \\
\rm{K_a \gamma_t(H-y) } & \rm{\mbox{if } WL< y < H }
\end{array}
\right.
\\
&\rm{ p_{ey}} =
\left\{
\begin{array}{ll}
\rm{\big[\gamma_t(H-WL) + (\gamma_t - \gamma_w)(WL-y)\big] } & \rm{\mbox{if } y \leq WL} \\
\rm{ \gamma_t(H-y) } & \rm{\mbox{if } WL< y < H }
\end{array}
\right.
\\
& \rm{ p_{e}} = \sqrt{  \rm{ p_{ex}}^2 + \rm{ p_{ey}}^2    }
\end{align}
where $\rm{WL = 7 m}$ is the water level upstream of the dam and $\rm{H = 10m}$ the dam's height. Specific weights $\rm{\gamma_w,\gamma_t}$ correspond to water and tailings material respectively, $\rm{K_a = \frac{1-sin(\phi)}{1+sin(\phi)} }$, is the active earth pressure coefficient, $\phi$ being the tailings' angle of friction, and $\rm{x,y}$ denote the horizontal and vertical directions.   \par
\begin{figure}[H] 
	\centering
	\begin{subfigure}[H]{0.65\columnwidth}
		\includegraphics[width=1.2\columnwidth, trim={1cm 0cm  0 0cm}]{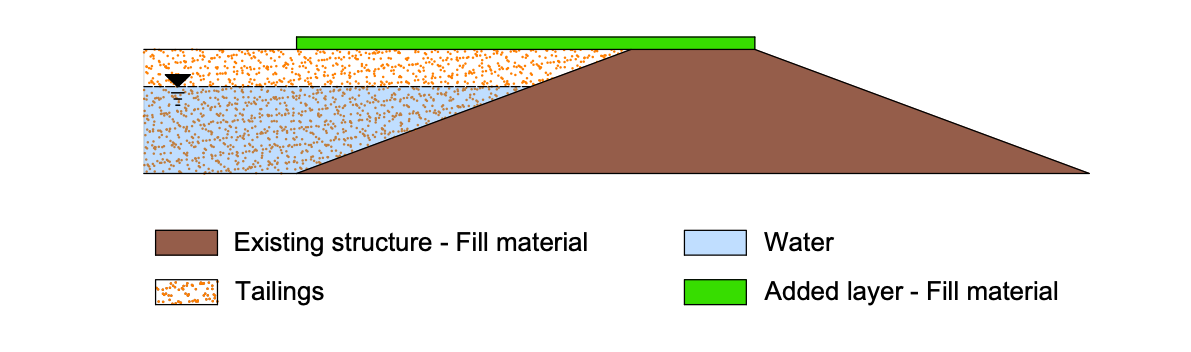}
		\caption{}

	\end{subfigure}
	
	\begin{subfigure}[H]{0.65\columnwidth}
		\includegraphics[width=1.2\columnwidth, trim={1cm 0cm  1cm 0cm}]{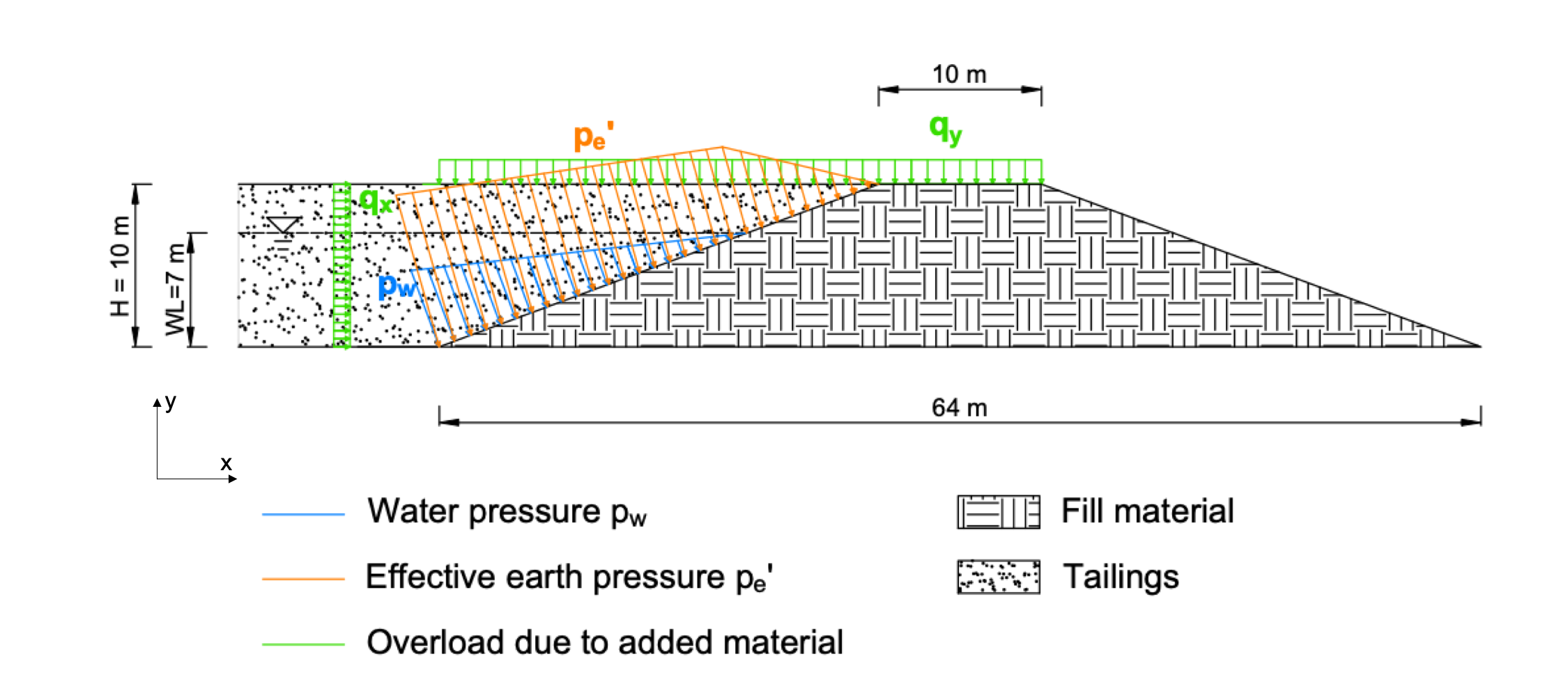}
		\caption{}

	\end{subfigure}
	
	\caption{Final conditions. (a) Conceptual sketch (b) Loads due to water and tailings as modeled }
 	\label{Final Conditions}
\end{figure}

Following, a load that corresponds to a level raise by $\rm{1 m}$ is gradually applied to the top of the structure. The load increases over a duration of 10 days and is then kept constant for the rest of the simulation. This setup is meant to simulate realistic conditions, that is, a small raise, no more than a meter per year, that is preceded and followed by a period of approximately equal time, during which no tailings are deposited, and the fill material is given some time to consolidate. The water table upstream, remains stable throughout the simulation. The final conditions are displayed in Figure \ref{Final Conditions}.

The loads attributed to the 1 meter thick layer of added material are estimated as,
\begin{align}
\rm{q_x= K_a \times q_y}\\
\rm{q_y=\gamma_f \times 1 m}
\end{align}
where $\rm{\gamma_f}$ is the added fill material specific weight.\par
\begin{figure}[H]
	\centering
	\includegraphics[width=0.7\columnwidth, trim={0 0cm  0 0cm}]{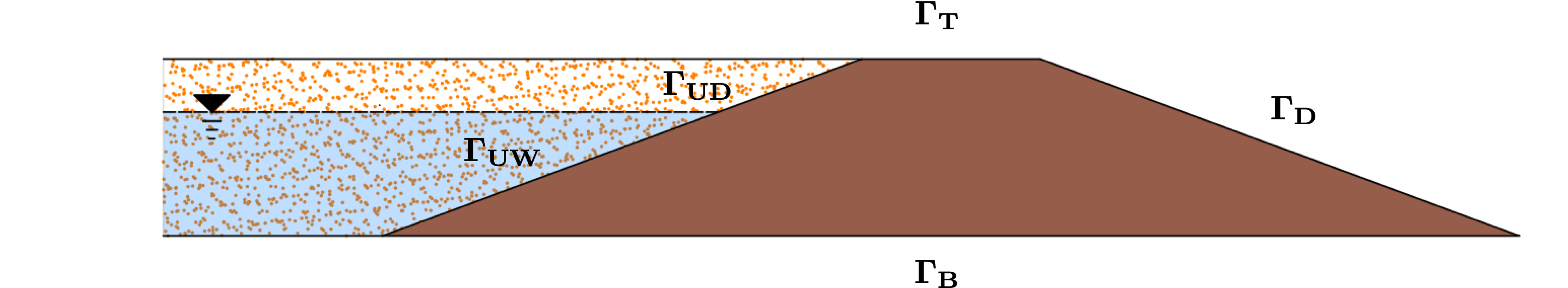}
	\caption{Boundary parts: Upstream Dry, Upstream Wet, Bottom, Downstream, Top}
	\label{Boundaries}
\end{figure}

Therefore the conditions that bound the problem are of type Dirichlet, Newman and Robin. In Figure \ref{Boundaries}, 5 boundary parts are defined as $\rm{\Gamma_{UD}, \Gamma_{UW}, \Gamma_{B},\Gamma_D, \Gamma_T}$ that denote Upstream Dry (above water table), Upstream Wet, Bottom, Downstream and Top boundary parts respectively. 
The boundary conditions introduced to the Finite Element model are written as
\begin{align}
\rm{\mathbf{u}_x} =\rm{ \mathbf{u}_y = 0 } \quad &\text{on} \ \mathbf{ \Gamma_B} \\
\rm{\boldsymbol{\sigma}\cdot \mathbf{n} } =\rm{ \begin{bmatrix} \rm{q_x+p_{ex}} & \rm{q_y + p_{ey}} \end{bmatrix}^\intercal } \quad &\text{on} \ \mathbf{ \Gamma_{UD}} \\
\rm{\boldsymbol{\sigma}\cdot \mathbf{n} } =\rm{ \begin{bmatrix} \rm{q_x+ p_{wx}+p_{ex} } & \rm{q_y+ p_{wy}+p_{ey}} \end{bmatrix}^\intercal } \quad &\text{on} \  \mathbf{ \Gamma_{UW} }\\
\rm{\boldsymbol{\sigma}\cdot \mathbf{n} } =\rm{ \begin{bmatrix} \rm{0 } & \rm{q_y} \end{bmatrix}^\intercal } \quad &\text{on} \  \mathbf{ \Gamma_{T}}\\
\rm{p = \gamma_w \times (WL-y)} \quad &\text{on} \ \mathbf{ \Gamma_{UW}} \\
\rm{ \mathbf{q} \cdot \mathbf{n}} =\rm{0} \quad &\text{on} \ \mathbf{ \Gamma_{UD}}\cup  \mathbf{ \Gamma_{B}}\cup  \mathbf{ \Gamma_{T}}\\
\rm{ \mathbf{q}\cdot \mathbf{n} } =\rm{\langle \beta p \rangle} \quad &\text{on} \ \mathbf{ \Gamma_{D}}.
\end{align}
The relations among the boundary parts definitions given here, to the ones given in equations \eqref{BC 1} - \eqref{BC 2} and \eqref{BC 3} - \eqref{BC 5} may be written  $\rm{\Gamma_D^u} =  \mathbf{ \Gamma_{B}}$, $\rm{\Gamma_N^u} = \mathbf{ \Gamma_{UD}}\cup  \mathbf{ \Gamma_{UW}}\cup  \mathbf{ \Gamma_{T}}$,  $\rm{\Gamma_D^p} =  \mathbf{ \Gamma_{UW}}$, $\rm{\Gamma_N^p} =  \mathbf{ \Gamma_{UD}}\cup  \mathbf{ \Gamma_{B}}\cup  \mathbf{ \Gamma_{T}}$, $\rm{\Gamma_R^p} = \mathbf{ \Gamma_{D}}$.

\subsection{Implementation: Full and reduced order solvers}
A FEM code for the problem stated above was developed in the FEniCS open-source platform. Following, the Reduced Basis method was used to create a low-order solver for the parametrized problem. \par 

The values of the parameters used in the model are given in Tables \ref{Physical Parameters} and \ref{Numerical Parameters}. The values were chosen such that they fall into ranges that are usually observed in tailings dams  \citep{bhanbhro_mechanical_2014} \citep{qiu_laboratory_2001}. 
\begin{table}[H] 
	\begin{center}
		\begin{tabular}{l | c | c | c }
			\hline \hline
			Parameter & Symbol & Units & Value\\
			\hline \hline
			Gravitational acceleration & $\rm{g}$ & $ \rm{m/s}$ & 10 \\
			Water bulk modulus & $\rm{K_w} $ & MPa & $2.2\times 10^{3}$\\
			Specific weight of water& $\rm{\gamma_w}$ &  $\rm{kN/m^3}$ & $10$\\
			\hline
			\multicolumn{4}{c}{Embankment fill soil material}\\
			\hline
			Particle density & $\rm{\rho_s} $ &  $\rm{kg/m^3}$ & $2.7\times 10^3$\\
			Young's Modulus &$\rm{E}$& MPa & 40\\
			Poisson's ratio & $\nu$ & -& $0.3$\\
			Porosity & $\eta$ & -  &0.38\\
			\hline
			\multicolumn{4}{c}{Tailings and added fill material}\\
			\hline
			Added fill material specific weight& $\rm{\gamma_f}$ &  $\rm{kN/m^3}$ & $21$\\
			Tailings specific weight & $\rm{\gamma_t} $ &  $\rm{kN/m^3}$ & $21$\\
			Tailings friction angle &$\rm{\phi} $ &  $^\circ$ & $35$\\
			\hline
			\multicolumn{4}{c}{Van Genuchten Model \citep{van_genuchten_closed-form_1980}}\\
			\hline
			Saturated VWC & $\rm{\theta_s}$ &- & 0.38\\
			Residual VWC & $\rm{\theta_r}$ &- & 0.038\\
			Parameter ($\approx$ inverse of air entry suction head) &$\alpha$ & $\rm{m^{-1}}$ & 0.1\\
			Fitting Parameter & $\rm{m}$ &  - & 0.184 \\
			Saturated hydraulic conductivity & $\rm{k_s}$ & $ \rm{m/s}$ &  $\rm{[10^{-9},...10^{-7}]}$
	\end{tabular}  \end{center}
	\caption {Values of physical parameters used in the model}
	\label{Physical Parameters}
\end{table}

\begin{table}[H]	
	\begin{center}
		\begin{tabular}{l  | c  }
			\hline
			\multicolumn{2}{c}{Discretization}\\
			\hline
			Mesh &  1382 elements, 774 nodes, unstructured\\
			FE Displacement & P2\\
			FE Pressure & P1 \\
			\hline
			\multicolumn{2}{c}{Time integration: $\theta$-scheme}\\
			\hline
			$\theta$& 0.75 \\
			Time step $\rm{dt}$ & 0.1 days \\		
	\end{tabular} \end{center} 
	\caption {Numerical parameters used in the model in the 2D scheme}
	\label{Numerical Parameters}
\end{table}

The parameter chosen to be examined is the material saturated hydraulic conductivity $\rm{k_s}$. As it has been mentioned above, hydraulic properties of the materials that exist in tailings dams feature high uncertainty and may vary in time. 
The parametric domain is taken such that the extreme values are realistic in the framework of tailings dams. The saturated hydraulic conductivity of the fill material takes values in $\rm{[10^{-9},...10^{-7}](m/s)}$. \par
\subsubsection*{Solving the loading problem with FE}
\begin{figure}[H] 
	\centering
	\begin{subfigure}[H]{0.6\columnwidth}
		\includegraphics[width=1\linewidth]{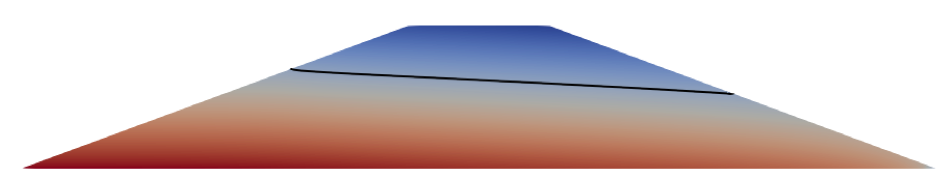}
		\caption{}
		\label{fig:Ng1} 
	\end{subfigure}
	
	\begin{subfigure}[H]{0.6\columnwidth}
		\includegraphics[width=1\linewidth]{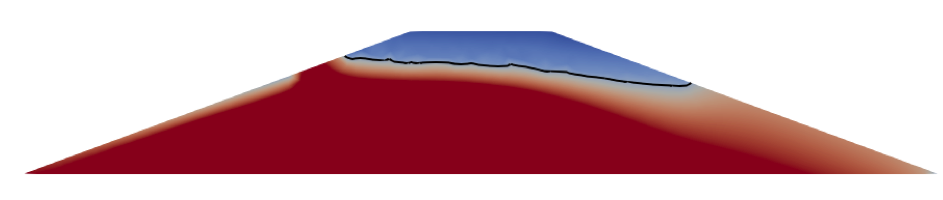}
		\caption{}
		\label{fig:Ng2}
	\end{subfigure}
	
	\begin{subfigure}[H]{0.6\columnwidth}
		\includegraphics[width=1\linewidth]{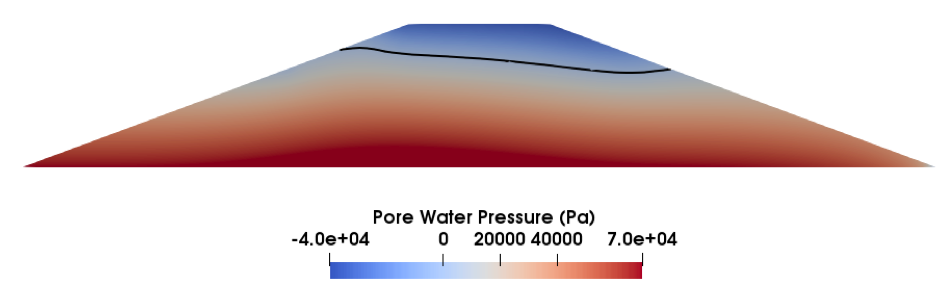}
		\caption{}
		\label{fig:Ng3}
	\end{subfigure}
	
	\caption[Pore Water Pressure field at time instances]{(a) Initial conditions. Steady state with fixed upstream water table at 7 m (b) At $\rm{t = 10}$ days, after 10 days of loading. The maximal overpressure is reached. At this point, the loading stops and pressure starts dissipating. (c) At $\rm{t = 60,6}$ days, end of the simulation. The black line indicates the position of the phreatic line.}
\end{figure}

In Figures \ref{fig:Ng1}-\ref{fig:Ng3} the pore water pressure fields acquired by the FE model, solving for $\rm{k_s=10^{-9} m/s}$, for 3 different time instances are shown. Figure \ref{fig:Ng1} corresponds to the initial state before the loading is applied. Figure \ref{fig:Ng2} represents the pore pressure state after 10 days of loading. During the loading time, changes in the displacement field reflect settlement in the dam due to overload. The hydro-mechanical coupling induces an increase of pore pressure, as the material is compressed, and the pore space reduced. Overpressure has been built due to the weight of the added layer of material and the water table has risen. After this point the load ceases to increase and the domain undergoes consolidation as the water table falls. The simulation stops when a steady state is reached. The pressure field is considered to be in steady conditions when the $L_2$ norm of the difference between the fields corresponding to two subsequent time steps, is smaller than $\rm{10^{-2}}$. For $\rm{k_s=10^{-9} m/s}$ this occurs at $\rm{t=60,6}$ days and the final solution field can be visualized in Figure \ref{fig:Ng3}.

\subsubsection*{Setting up the ROM: Offline stage}
As explained in section \ref{Methodology}, the Reduced Basis is constructed by means of sampling the high-fidelity solution manifold, that is, in this case, the set of solutions obtained by the full-order Finite Element solver. \par
Specifically, the full order problem was solved for $\rm{k_s =[1,3,5,7,10,30,50,70,100]\times 10^{-9}}$, and the solutions were stored in two separate snapshot matrices as in equations \eqref{Mu} and \eqref{Mp}. Each of the snapshots have a duration of 10 days, during which the load is applied, plus the time that is needed for steady state conditions to be reached. The time required for steady state conditions to be reached after loading depends on the hydraulic conductivity of the material. The built-up pressure requires longer time to dissipate in a less permeable material. Thus the snapshots have different durations. \par
\begin{figure}[H]
	\centering
	\includegraphics[width=500pt, trim={0 1cm  0 1cm}]{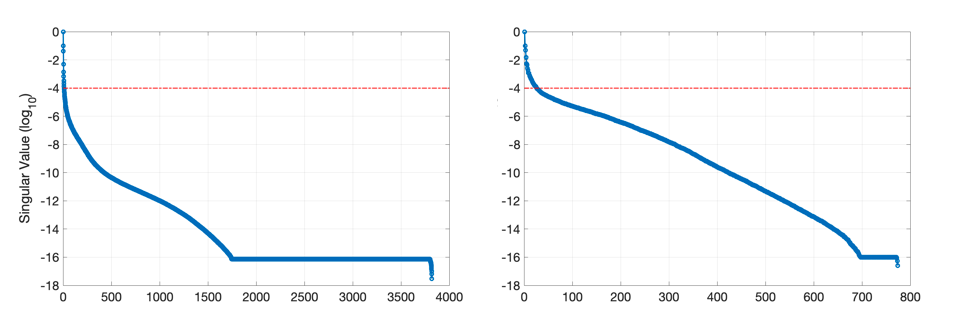}
	\caption{Left singular matrix truncation. Left: Displacement. Right: Pressure}
	\label{Truncation}
\end{figure}

Singular value decomposition was applied to the two matrices resulting to the left singular matrix, that was truncated to yield the Reduced Basis.
The truncation criterion is based on the singular values. In Figure \ref{Truncation} the singular values that correspond to each of the vectors of the left singular matrix for the two fields are plotted. The y-axis is in logarithmic scale and it is normalized with respect to the first -and largest- singular value. The values drop rapidly in both cases. The first vectors contribute significantly to the description of the solution set, and must, therefore, be included to the Reduced Basis, while as the singular value decreases, the corresponding vectors convey less information about the data, that is, the snapshot matrix \par
 The red dashed line denotes the truncation threshold. In this case, vectors that correspond to singular values that are smaller than the first one by 4 orders of magnitude or more, are discarded. Obtaining a ROM with an accuracy higher than that, would be beyond the scope of the ROM, since the accuracy in the problem is limited by that of the sensor measurements. \par
This truncation threshold yields a basis that is comprised of just 9 vectors for displacement and 25 for pressure, thus 34 is the size of the reduced system of equations. To put this number of reduced unknowns in perspective, the high-fidelity problem dimension, related to mesh resolution and polynomial degree is 6632 degrees of freedom.

\subsubsection*{Online stage: Results and comparison}
\begin{figure}[H]
	\centering
	\includegraphics[width=\linewidth]{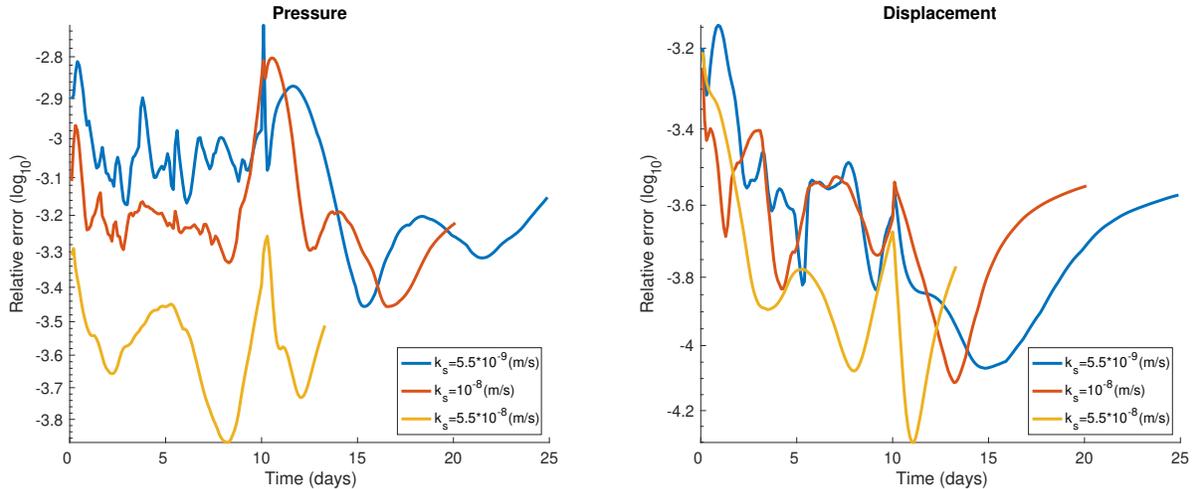}
	\caption{Relative error of low-order approximation of the pressure field with respect to full order solution over time, estimated over the entire domain, for 3 different parametric values. Left: Pressure field. Right: Displacement field.}
	\label{2D_Errors}
\end{figure}

Having populated the transformation matrices $\mathbf{B_u}$ and $\mathbf{B_p}$, the problem may now be solved, for any parameter within the examined range, by assembling the system of equations and projecting it to the reduced space in which the approximation will be sought, as shown in equation \eqref{RB System}.
In Figure \ref{2D_Errors}  the relative error of the low order approximation with respect to the high-fidelity solution, is plotted over time. The error is estimated as $ \rm{e =\abs{ \frac{X_{RB}- X_{FEM}}{X_{FEM}}}}$, where $\rm{X_{RB}}$ and $\rm{X_{FEM}}$ represent the approximation and the high-fidelity solution respectively.\par
Of the three values examined in Figure \ref{2D_Errors}, one is a snapshot value, namely $\rm{k_s= 10^{-8}} \ m/s$, and the other two are values that were not sampled. 
The errors for both fields  and for all parametric values remain quite low, despite the small number of base vectors used. In fact, the error is much lower than the typical accuracy of measurement of the instrumentation that corresponds to the quantities evaluated. Note that the error doesn't seem to be significantly smaller for the case of the snapshot value $\rm{k_s= 10^{-8}} \ m/s$. This indicates that the Reduced Basis sufficiently describes the solution states that correspond to the entire spectrum of parametric values. Moreover, it is worth mentioning that the reduced order model yields a small error even for the parametric value that corresponds to a snapshot, which is an expected behavior, considering that the snapshot matrix was truncated after the singular value decomposition.  \par
The ROM runs 3 times faster, by average, than the full-order model. That is a significant boost of computational efficiency, especially considering the relatively low mesh resolution. \par 
To test the effect of mesh resolution to the computational time reduction, another scheme was run, for the same problem, using a denser mesh. In this scheme the high-fidelity system of equations has 2975 degrees of freedom for pressure and 23164 for displacement. The same number of snapshots were taken and corresponded to the same parametric values. The reduced bases that yield the same level of accuracy as the first scheme have 23 base vectors for the pressure field and 9 for the displacement field. The number of vectors needed is very close to the previous scheme, despite the large difference in the size of the high-fidelity problem. In other words, the size of the reduced scheme is decoupled from the size of the high-fidelity scheme \citep{hesthaven_certified_2016}. This is an essential advantage of the Reduced Basis method. 
It is worth noting that in the case of non-linear problems, like the present, the reduced scheme cannot be fully decoupled from the FEM scheme. Due to the state-dependence of the FEM operators in the linearized system, they have to be reassembled in every iteration of the linearization scheme. The full order system has to be assembled and then transformed into the reduced order one. The assembly of these operators is related to the high-fidelity dimension. Therefore the efficiency gains are bounded. This issue is often addressed by the empirical interpolation methods \citep{hesthaven_certified_2016}\citep{quarteroni_reduced_2016}, the implementation of which, was considered outside the scope of this paper.

\subsection{Extension to 3D problems}
Literature suggests that 2 dimensional models cannot reflect the complex and variating seepage field \citep{lyu_comprehensive_2019}. Thus, 3 dimensional models have been proposed for the stability study of tailings dams \citep{lu2006three}, \citep{zhang_numerical_2020}. In this section, a seepage problem, similar to the 2 dimensional one above, will be solved in a 3D setting and a ROM will be developed and evaluated with respect to its accuracy and computational efficiency against the high-fidelity model.\par
The new geometry approximates an embankment constructed in a narrow, steep sided valley. This is a generic, invented geometry, meant to simulate common conditions in the construction of embankment dams. \par
\begin{figure}[H]
	\begin{subfigure}{.5\textwidth}
		\centering
		\includegraphics[width=.8\linewidth, trim={0 1cm  2cm 0}]{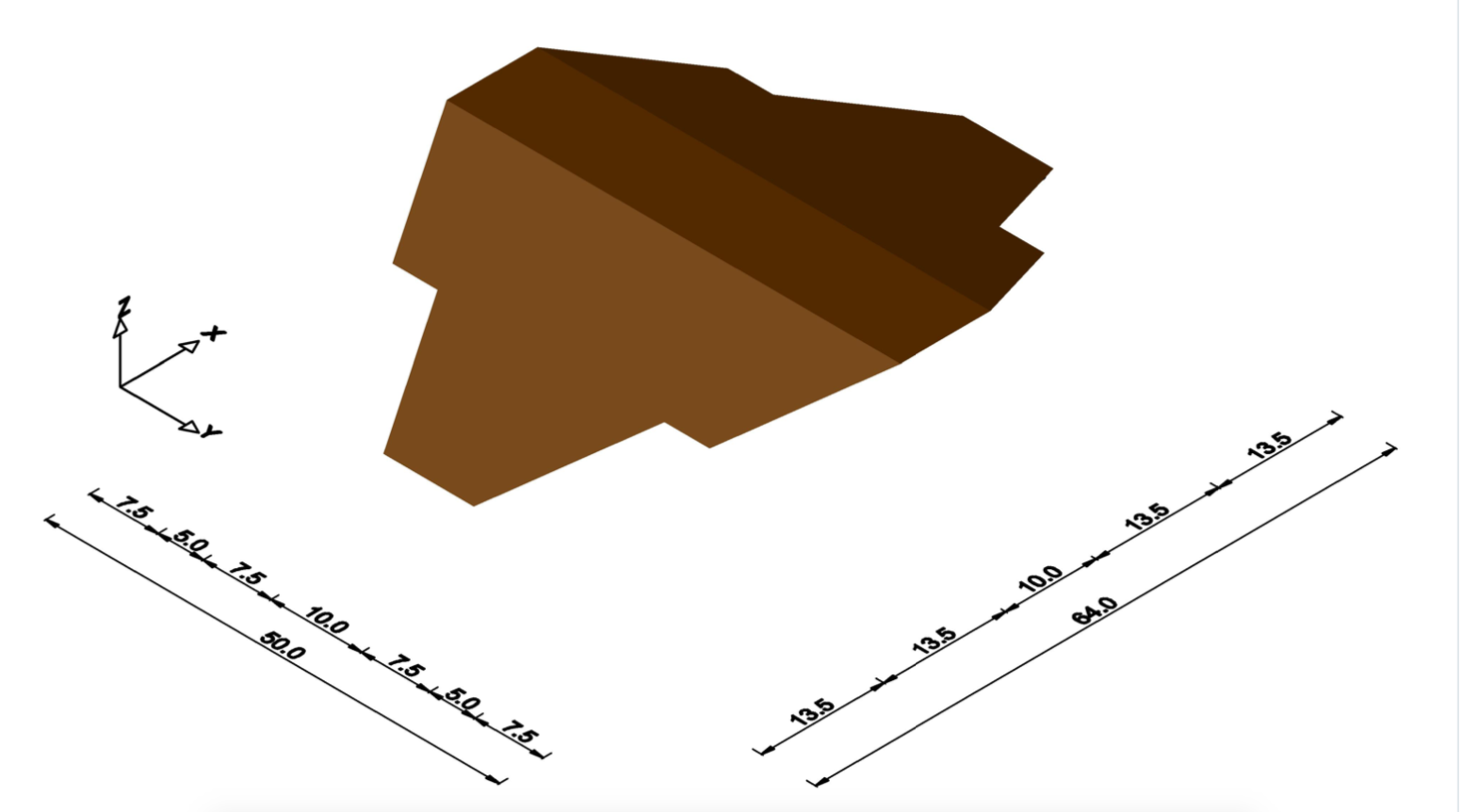}  
		\caption{3D Isometric view}
		\label{3D view}
	\end{subfigure}
	\begin{subfigure}{.5\textwidth}
		\centering
		\includegraphics[width=.8\linewidth]{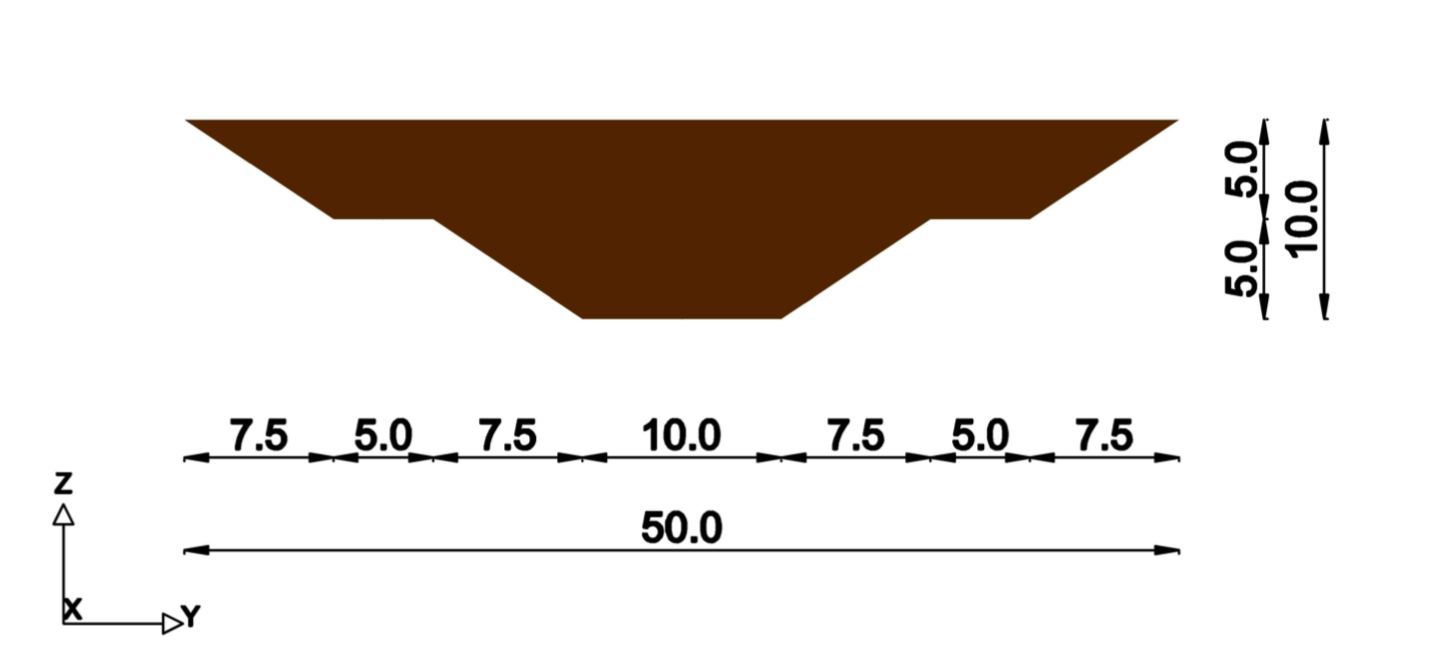}  
		\caption{Front view}
		\label{Front view}
	\end{subfigure}
\\	
	\begin{subfigure}{.5\textwidth}
		\centering
		\includegraphics[width=.8\linewidth]{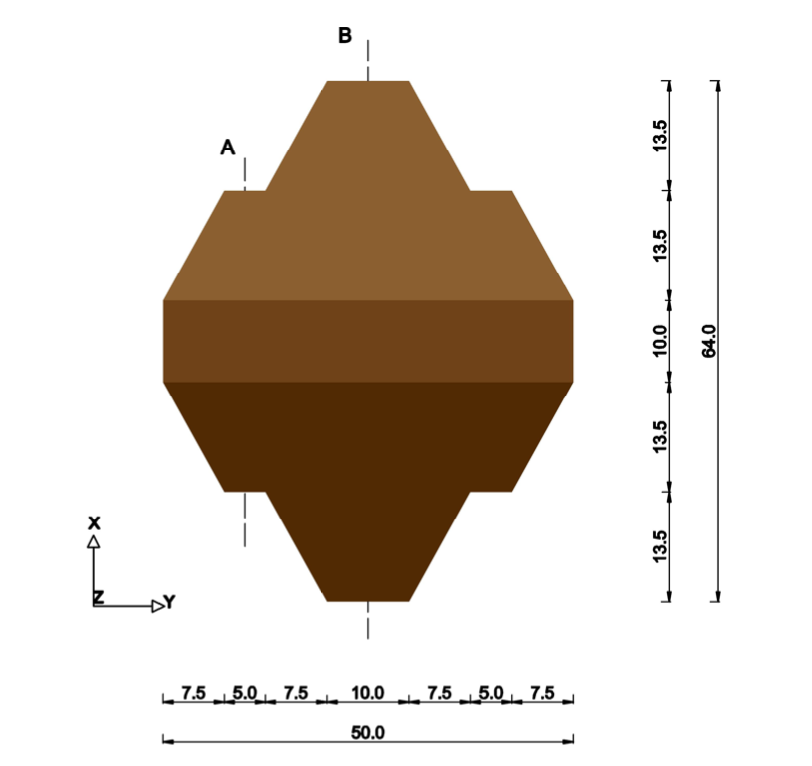}  
		\caption{Top view}
		\label{Top view}
	\end{subfigure}
	\begin{subfigure}{.5\textwidth}
		\centering
		\includegraphics[width=.8\linewidth]{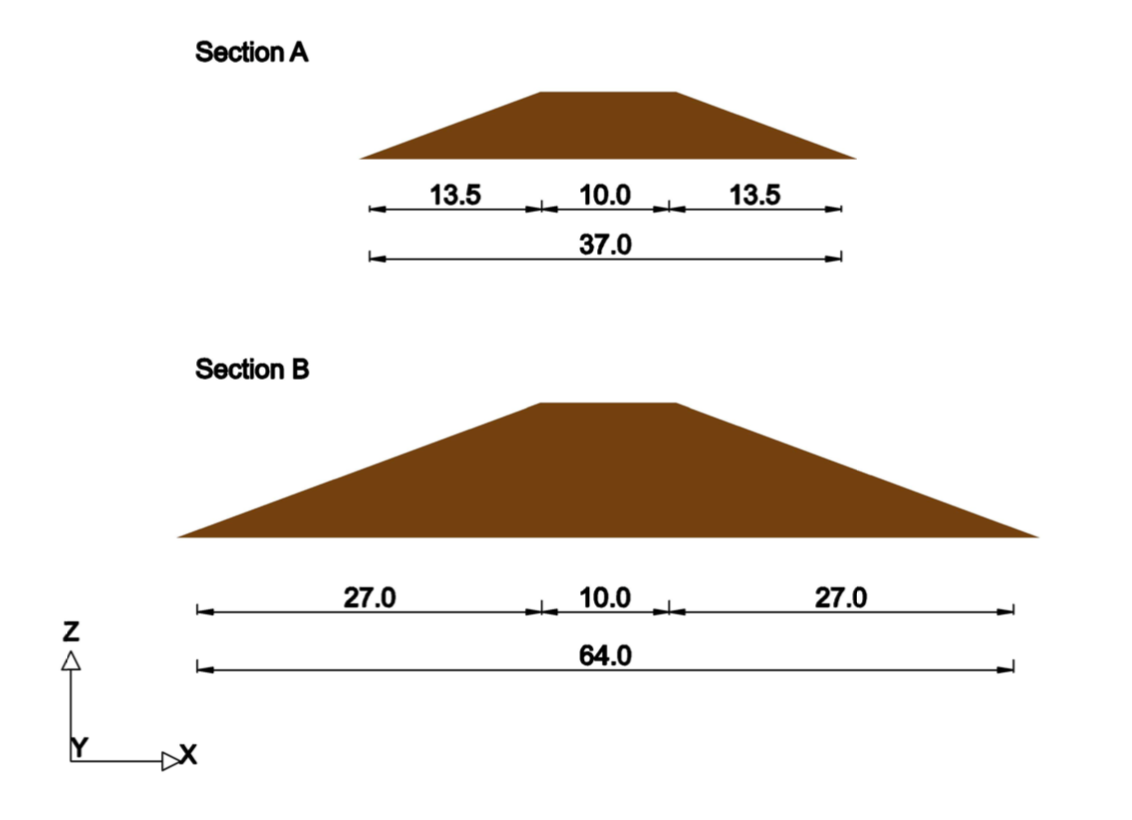}  
		\caption{Cross sections Top: Section plane A Bottom: Section plane B}
		\label{Cross sections}
	\end{subfigure}
	\caption{Views and cross sections of the 3 dimensional domain. The geometry assumes an embankment that has been constructed in a narrow valley, and has its foundations on two side slopes at the two extremes in the direction of Y axis. The side slopes are not displayed. }
	\label{Domain Views}
\end{figure}

 
In Figure \ref{Domain Views} the Y axis indicates the third dimension, while axes X and Z correspond to the dimensions that were considered in the plane strain approximation of the previously solved problem. The domain has a variating cross-section along the Y axis.  The two-dimensional domain of the previous setting corresponds to the middle section of the dam, ie the part that has its foundation on the valley (Figure \ref{Cross sections} Bottom). The parts that are founded on the lateral slopes, have smaller cross sections (Figure \ref{Cross sections} Top). The side slopes have been assumed excavated in two levels, creating two 5 m-high slopes of inclination 1,5:1.  
The domain is a 50 m-long prism, has 2 axes of symmetry, and an identical cross-section as the 2D domain shown in Figure \ref{Initial Conditions pic}. \par 

The 3D tetrahedral mesh was created in Gmsh open-source mesh generating software \citep{geuzaine_gmsh_2009}. The mesh is made out of 41477 tetrahedral and triangular cells and 9896 nodes. Using the Taylor-Hood (P2-P1) element for the description of the displacement and pressure fields as before, the resulting system is comprised of 202179 and 9896 degrees of freedom for the two fields respectively. \par 
As for the problem setup, it remains similar to the 2D case with some alterations. The initial condition is evaluated by solving the steady state problem for an upstream water level at 7 m. The added material is now placed in two layers of 50 cm each. The first layer is deposited in the first 5 days, gradually, across the Y axis. Then the second 50 cm layer is added on top, during days 5-10, again, gradually across the Y axis. After the period of 10 days, the full load that represents the 1m-thick layer remains constant. The bottom boundary is mechanically constrained. The parameters used have the values listed in Table \ref{Physical Parameters}.  During the level raise phase in a tailings dam's life,  thin layers of fill material are deposited along the length of the structure. The load increases gradually along the 3rd axis of the dam, i.e. the dimension that is not considered in a 2D plane strain approximation. The problem simulates conditions that may occur in the context of an actual tailings dam and cannot be sufficiently approximated in a 2D plane strain setup.\par
\begin{figure}[H]
		\centering
	\begin{subfigure}{.65\textwidth}
		\centering
		\includegraphics[width=\linewidth, trim={1cm 1cm  1cm 0}]{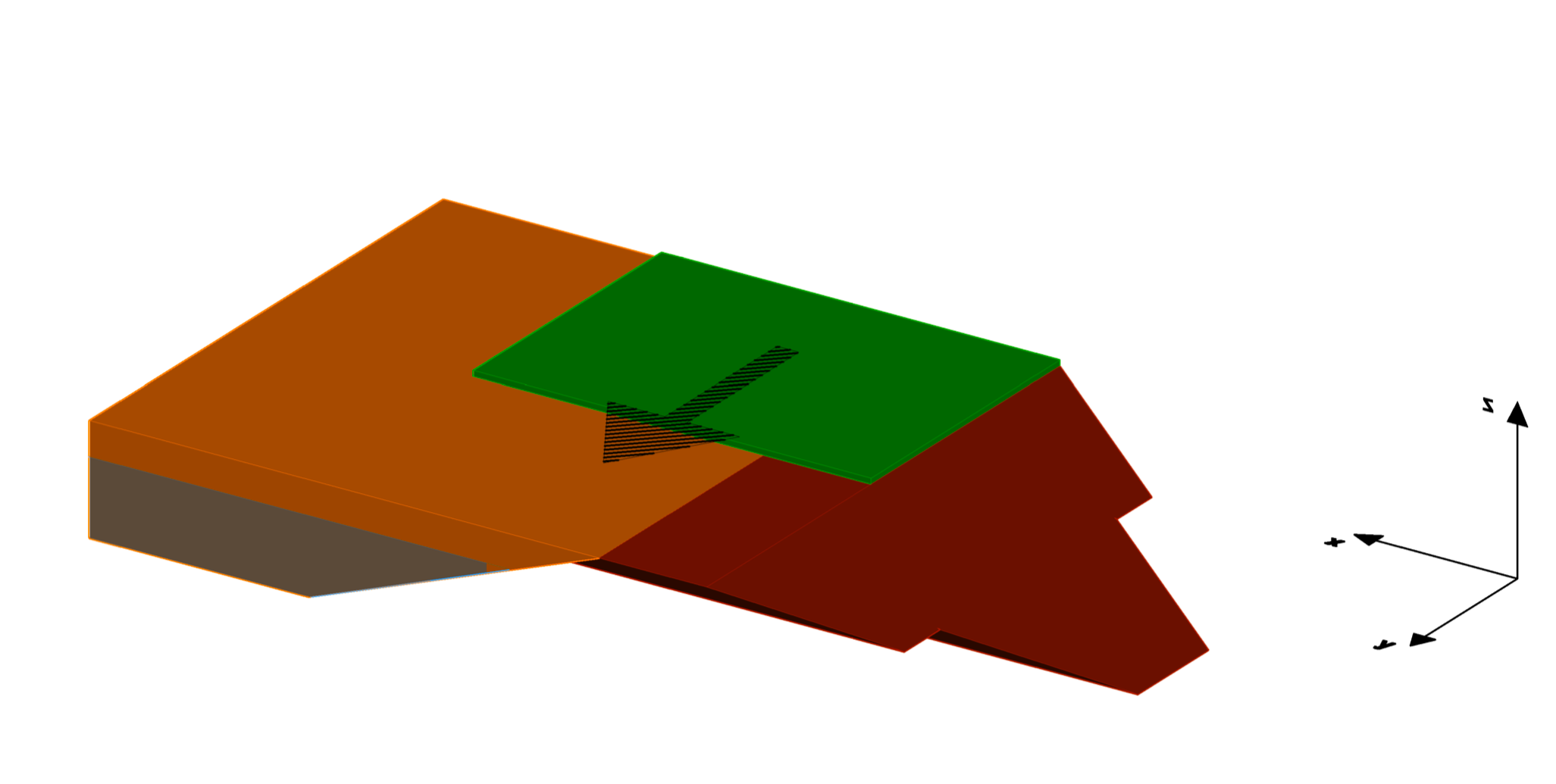}  
		\caption{After 2,5 days.}
		\label{1st_layer}
	\end{subfigure} \\
	\begin{subfigure}{.65\textwidth}
		\centering
		\includegraphics[width=\linewidth, trim={1cm 1cm  1cm 0}]{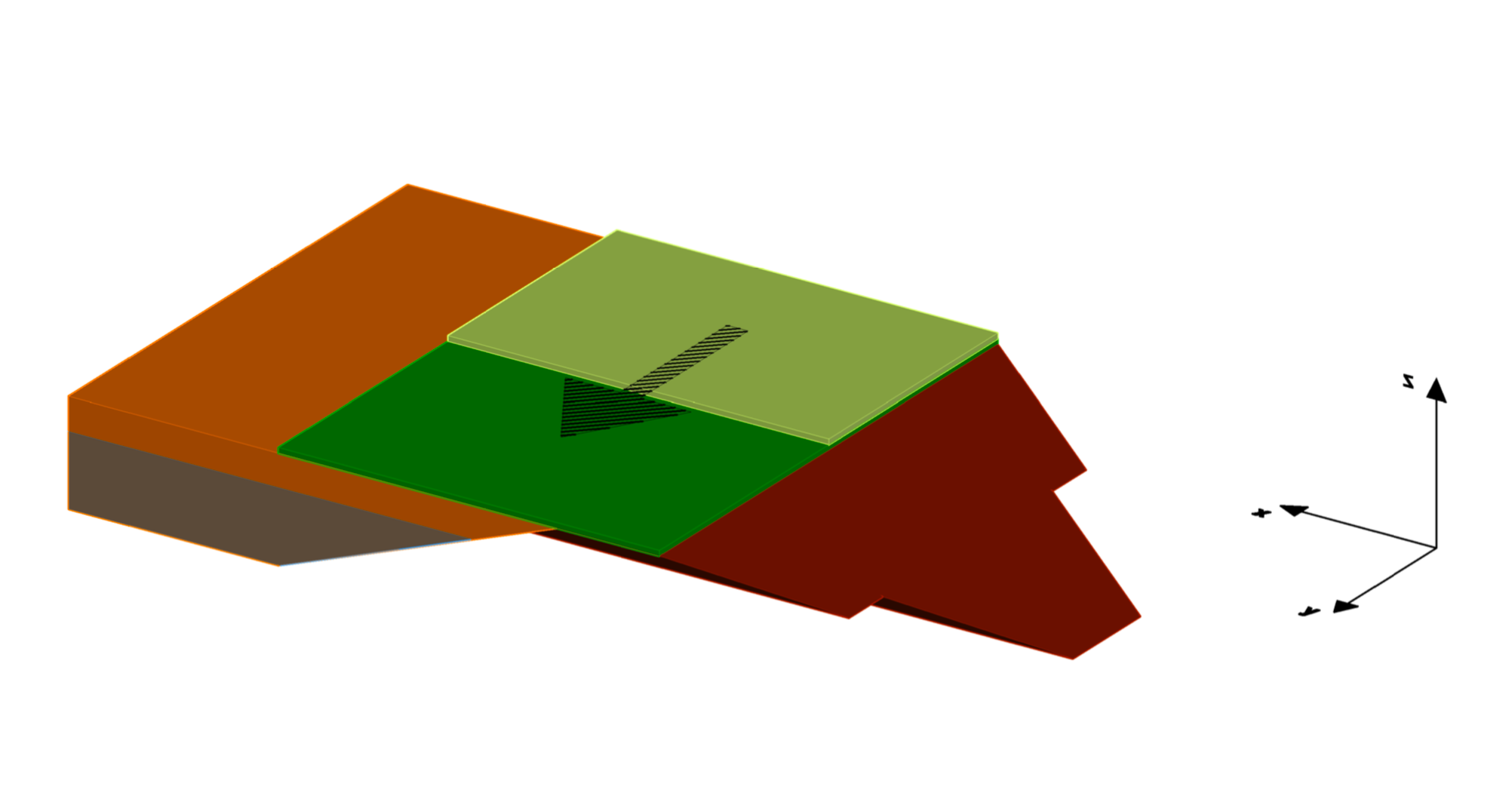}  
		\caption{After 7,5 days}
		\label{2nd_layer}
	\end{subfigure}

	\caption{Deposition of the two 50 cm - thick layers of fill material on the top and upstream side of the structure. The first layer is shown in dark green color and the second in light green color. After 2,5 days of loading half of the first layer has been deposited. After 7,5 days the first layer and half of the second layer have been deposited. Upstream is in the direction of the x axis. Unsaturated tailings are depicted in orange, saturated tailings in gray. The lateral foundation slopes are not depicted. }
\end{figure}

\begin{figure}[H]
	\begin{subfigure}{\textwidth}
		\centering
		\includegraphics[width=.65\linewidth, trim={0 1cm  2cm 0}]{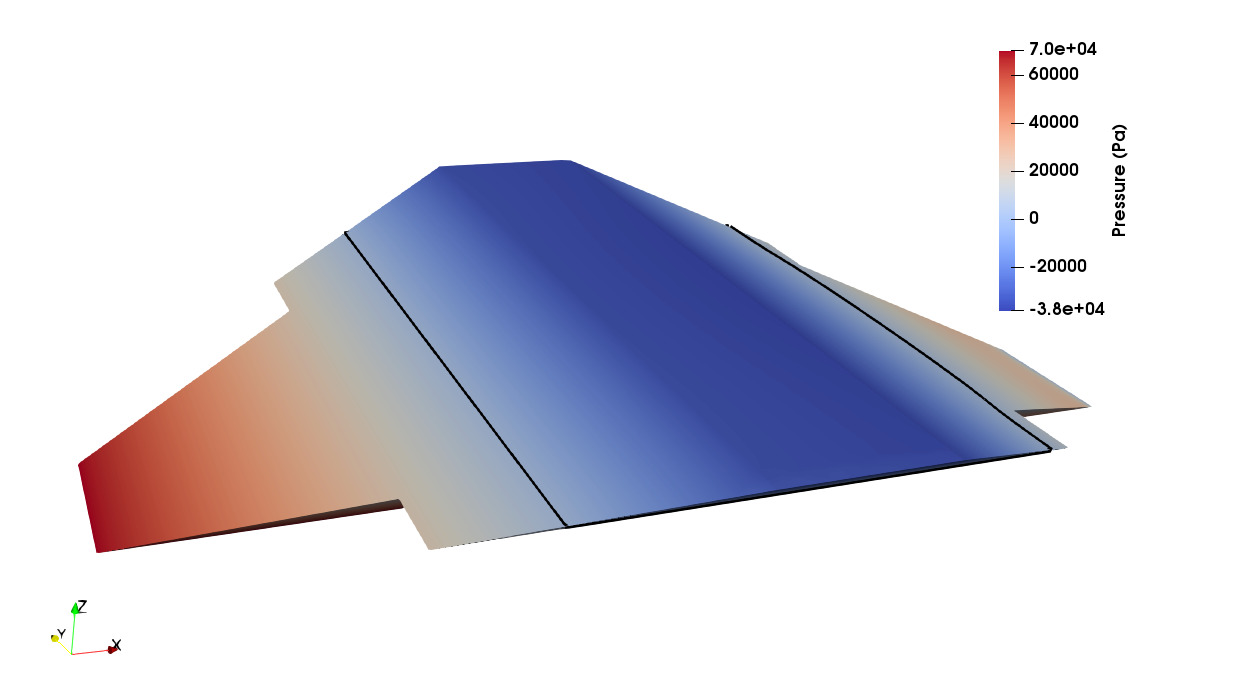}  
		\caption{t=0}
		\label{3D t0}
	\end{subfigure}\\
	\begin{subfigure}{\textwidth}
		\centering
		\includegraphics[width=.65 \linewidth]{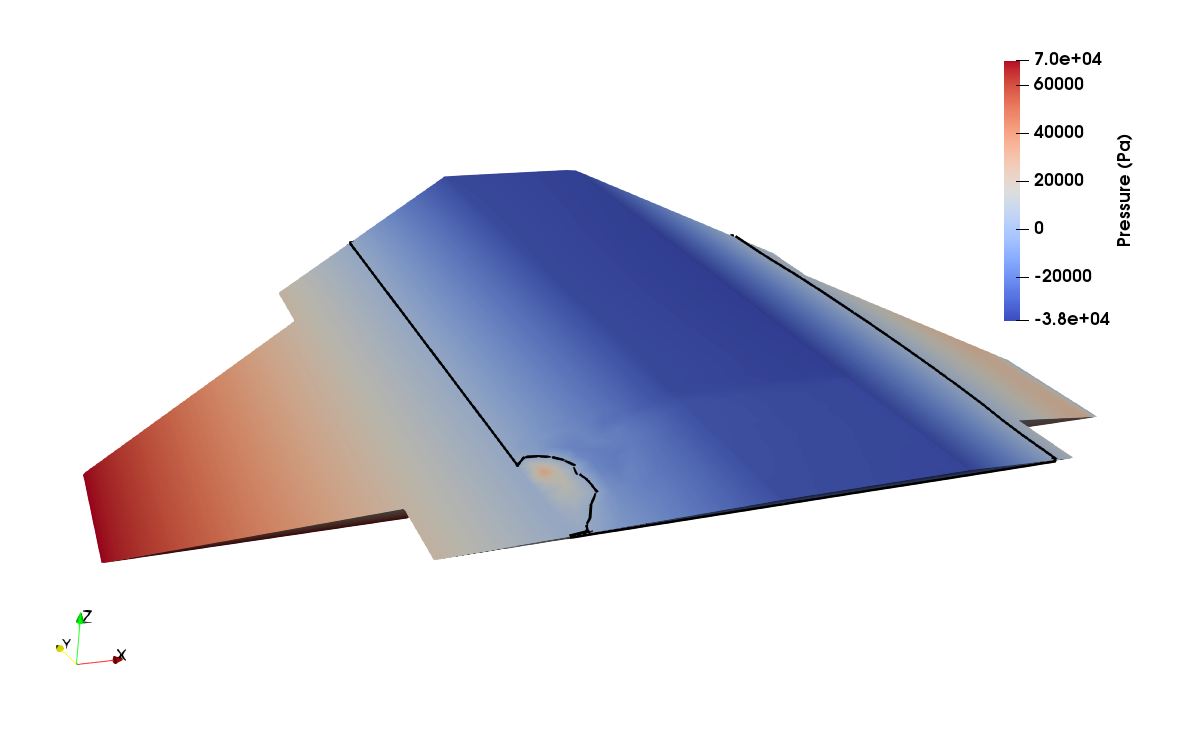}  
		\caption{t = 1 day}
		\label{3D t1}
	\end{subfigure}\\
	\begin{subfigure}{\textwidth}
	\centering
	\includegraphics[width=.65\linewidth]{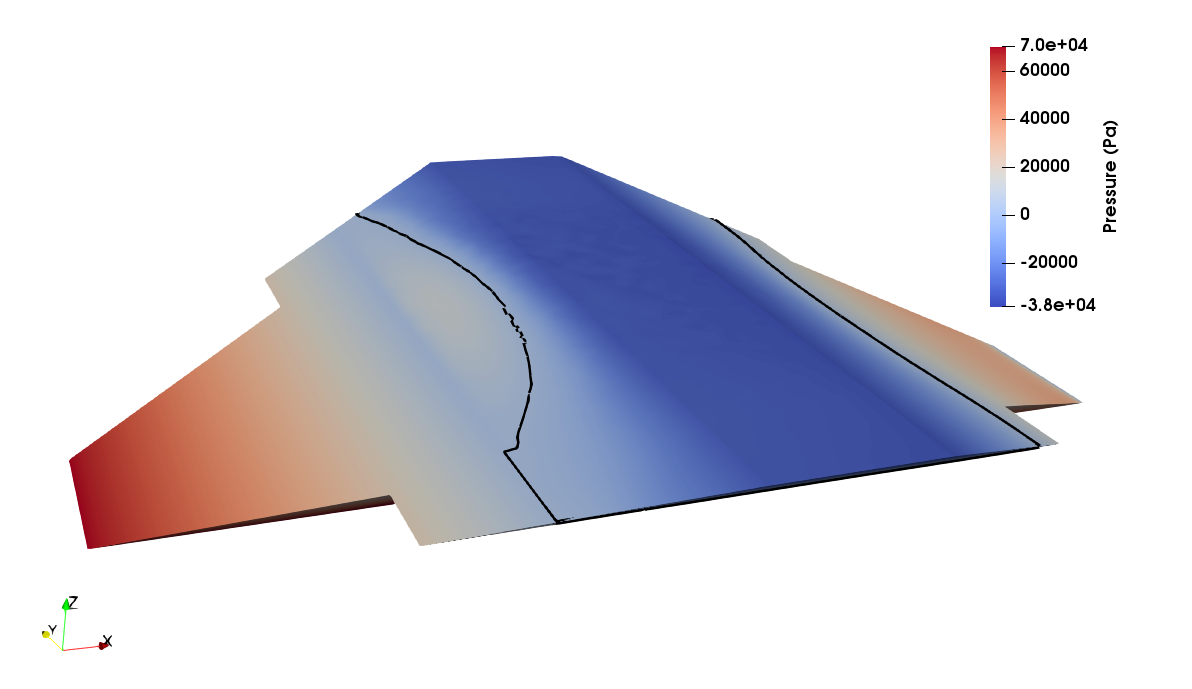}  
	\caption{t = 10 days}
	\label{3D t10}
\end{subfigure}
\caption{Pore water pressure distribution on different time instances (a) Initial conditions. Steady state condition with an upstream level at 7m. (b) After 1 day of loading. In some part of the domain overpressure has built up. (c) After 10 days of loading the top of the dam and impoundment. The reservoir is considered to be located on the left side of the figures, i.e. upstream is opposite to the x axis direction. }
\label{3D simulation}
\end{figure}

 
In Figures \ref{1st_layer} and \ref{2nd_layer} the loading is illustrated for clarity. Two layers of soil are placed on top of the dam, in order to raise the level of the dam by 1 meter in the upstream manner, as shown in Figure \ref{up_down_centre}.
In Figure \ref{3D simulation}, the effect that the gradual load application has on the pore pressure field is illustrated. As the load is applied from one extreme to the other in the y direction, the water table in the area rises to be then lowered after pressure is dissipated through consolidation. By the time the loading procedure is finished after 10 days, in some parts of the dam the water table has almost reached its final position after consolidation. 
The Reduced order scheme was created following the same steps as in the 2D case. Snapshots of the high-fidelity solution manifold were taken for parametric values $\rm{k_s =[1,3,5,7,10,30,50,70,100]\times 10^{-9}}$. As above, each snapshot has a duration of 10 days of loading, plus the amount of time needed for steady state to be reached in each case. The two solution fields - displacement and pressure- were stored in snapshot matrices $\mathbf{M_u}$ and $\mathbf{M}_{\rm{p}}$, arranged as before, i.e. snapshots comprised of all time steps stored in a serial manner in the columns of the matrices. Singular value decomposition was applied to each matrix separately. \par
Much like in the 2D case, the left singular matrix was truncated with a criterion related to the singular values, admitting only vectors that correspond to singular values up to 4 orders of magnitude smaller than the largest one. The truncation yielded reduced bases comprised of 80 base vectors for displacement and 174 for pressure. Therefore the system of equations to be solved now is of dimensionality 214, instead of the full order problem with 212075 degrees of freedom.\par
\begin{figure}[H]
	\centering
	\includegraphics[width=\linewidth]{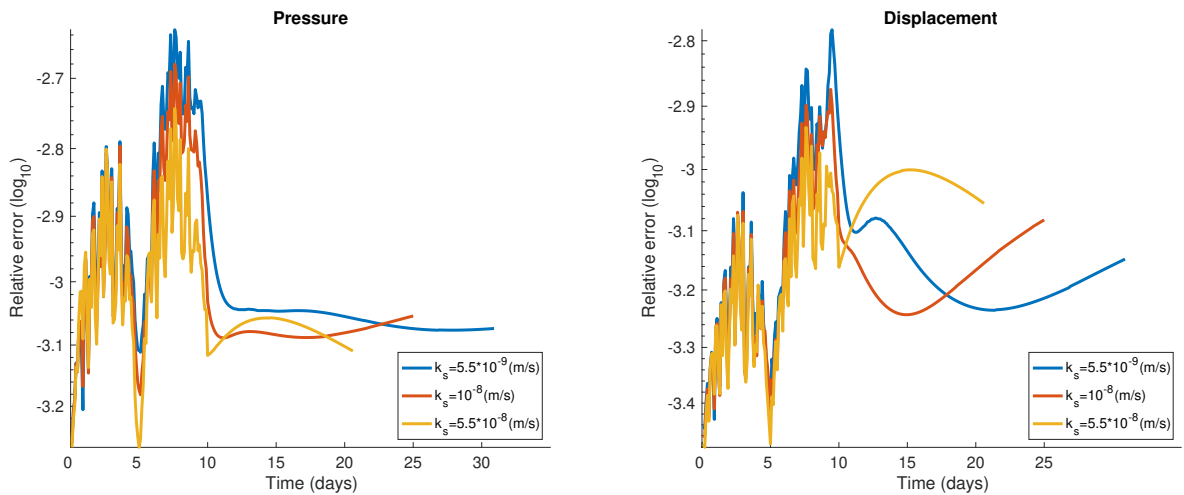}
	\caption{Relative error of low-order approximation of the pressure field with respect to full order solution over time, estimated over the entire 3D domain, for 3 different parametric values. Left: Pressure field. Right: Displacement field.}
	\label{3D_Errors}
\end{figure}

 
The same parametric values as before were examined in order to estimate the accuracy and computational efficiency gain of the RB approximation. In Figure \ref{3D_Errors} the $L_2$ norms of the relative error -estimated as the difference between full order and reduced order solutions- have been plotted for the two fields. 
In comparison to the 2D scheme, it seems that the ROM yields slightly lower accuracy, which remains however in the same order of magnitude. The accuracy can easily be improved by choosing a smaller truncation tolerance for the reduced basis.
For various parametric values that were tested, the ROM was found to run 8 - 15 times faster than the full-order model. This is a significant increase of gain in computational efficiency compared to the 2D case, which indicates that the time gain scales along with the size of the problem at hand.  \par 
\section{Conclusions and future work} \label{Conclusions}
In this work a low order model for the hydro-mechanically coupled problem has been developed, using the Reduced Basis technique. The POD-based method has been successfully applied. A first attempt to treat the parametrized system of PDEs and confirming a high level of accuracy has been demonstrated. The computational time needed is reduced in this simplified problem, while the accuracy of the solution is not compromised, in comparison to the full-order Finite Element solution. The method is promising, considering that the time reduction increases with increasing problem size. \par
It is worth noting that in the framework of tailings dams the usual practice is to model the retaining structure as well as the stored material, due to the fact that failure may originate in the impoundment. Sufficiently describing the pressure and displacement field -or other fields of interest- in such a large inhomogeneous domain can result in large systems, in which case the advantages of a reduced order model would be most prominent. \par  
 As it has been pointed out in this paper, computational efficiency is key for solving inverse problems, such as parameter identification in real-time, or near real-time. Enabling fast many-query solutions may contribute heavily to predictive monitoring applications in earthfil dams. On that note, the continuation of this work is likely to involve data assimilation applications and inverse problem solving for tailings dams.  \par
 The most straightforward step for the future of this work is the implementation of hyper-reduction techniques in order to treat the non-linear matrix and vector terms of the system, thus achieving higher computational efficiency. One quite promising method is the Discrete Empirical Interpolation (DEIM) proposed in \citep{chaturantabut_nonlinear_2010} and extended for treatment of matrix non-linear operators in \citep{negri_efficient_2015}. The method aims at identifying reduced approximations of nonlinear functions in a POD manner and has been successfully applied to coupled problems treated with FEM \citep{santo_hyper-reduced_2019}.    \par
 Another issue to be explored is high-dimensionality in the parametric domain. In tailings dams, parameters related to both the mechanical and hydraulic problems may feature high uncertainty and may fluctuate significantly as the structure evolves. Therefore, updating the model for multiple parameters as their values change is a relevant problem, and one that accentuates the need for order reduction in an inverse problem context. 

\section{Acknowledgements}
This work has received funding from the European Union’s Horizon 2020 research and innovation programme under the Marie Sklodowska-Curie grant agreement No 764636, for the research project ProTechTion. \par
We thank Professor Sebatian Olivella (Escola de Camins, Universitat Politècnica de Catalunya) for his much appreciated assistance. \par
Pedro Díez and Sergio Zlotink are grateful for the financial support provided by the Spanish Ministry of Economy and Competitiveness (Grant agreement No. DPI2017-85139-C2-2-R), by the Generalitat de Catalunya (Grant agreement No. 2017-SGR-1278), and by project H2020-RISE MATH-ROCKS GA no 777778. \par
\section*{References}
\bibliographystyle{humannat}

\bibliography{References}

\begin{thebibliography}{}

\bibitem[\protect\astroncite{Alnaes et~al.}{2015}]{aln_fenics_2015}
Alnaes, M.~S., B.~Kehlet, A.~Logg, C.~Richardson, J.~Ring, E.~Rognes, and G.~N.
  Wells\leavevmode\nopagebreak\newline 2015.
\newblock The {FEniCS} {Project} {Version} 1.5.
\newblock P.~~15.

\bibitem[\protect\astroncite{Alonso et~al.}{2005}]{alonso_review_2005}
Alonso, E.~E., S.~Olivella, and N.~M. Pinyol\leavevmode\nopagebreak\newline
  2005.
\newblock A review of {Beliche} {Dam}.
\newblock {\em Géotechnique}, 55(4):267--285.

\bibitem[\protect\astroncite{Badia et~al.}{2009}]{badia_coupling_2009}
Badia, S., A.~Quaini, and A.~Quarteroni\leavevmode\nopagebreak\newline 2009.
\newblock Coupling {Biot} and {Navier}–{Stokes} equations for modelling
  fluid–poroelastic media interaction.
\newblock {\em Journal of Computational Physics}, P.~~29.

\bibitem[\protect\astroncite{Bao et~al.}{2019}]{bao_data-driven_2019}
Bao, A., E.~Gildin, A.~Narasingam, and J.~S.
  Kwon\leavevmode\nopagebreak\newline 2019.
\newblock Data-{Driven} {Model} {Reduction} for {Coupled} {Flow} and
  {Geomechanics} {Based} on {DMD} {Methods}.
\newblock {\em Fluids}, 4(3):138.

\bibitem[\protect\astroncite{Bhanbhro}{2014}]{bhanbhro_mechanical_2014}
Bhanbhro, R.\leavevmode\nopagebreak\newline 2014.
\newblock Mechanical {Properties} of {Tailings}: {Basic} {Description} of a
  {Tailings} {Material} from {Sweden}.
\newblock Publisher: Unpublished.

\bibitem[\protect\astroncite{Chaturantabut and
  Sorensen}{2010}]{chaturantabut_nonlinear_2010}
Chaturantabut, S. and D.~C. Sorensen\leavevmode\nopagebreak\newline 2010.
\newblock Nonlinear {Model} {Reduction} via {Discrete} {Empirical}
  {Interpolation}.
\newblock {\em SIAM Journal on Scientific Computing}, 32(5):2737--2764.

\bibitem[\protect\astroncite{Clarkson et~al.}{2020}]{clarkson_real-time_2020}
Clarkson, L., D.~Williams, and J.~Seppälä\leavevmode\nopagebreak\newline
  2020.
\newblock Real-time monitoring of tailings dams.
\newblock {\em Georisk: Assessment and Management of Risk for Engineered
  Systems and Geohazards}, Pp.~ 1--15.

\bibitem[\protect\astroncite{Davies and Martin}{2002}]{davies_static_2002}
Davies, M. and T.~Martin\leavevmode\nopagebreak\newline 2002.
\newblock Static liquefaction of tailings–fundamentals and case histories.
\newblock In {\em Proceedings of {Tailings} {Dams} {ASDSO}/{USCOLD}}, Pp.~
  233--255, Las Vegas 2002.

\bibitem[\protect\astroncite{Esmaeili et~al.}{2020}]{esmaeili_generalized_2020}
Esmaeili, M., M.~Ahmadi, and A.~Kazemi\leavevmode\nopagebreak\newline 2020.
\newblock A generalized {DEIM} technique for model order reduction of porous
  media simulations in reservoir optimizations.
\newblock {\em Journal of Computational Physics}, 422:109769.

\bibitem[\protect\astroncite{Florentin and
  Díez}{2012}]{florentin_adaptive_2012}
Florentin, E. and P.~Díez\leavevmode\nopagebreak\newline 2012.
\newblock Adaptive reduced basis strategy based on goal oriented error
  assessment for stochastic problems.
\newblock {\em Comput. Methods Appl. Mech. Engrg.}, P.~~12.

\bibitem[\protect\astroncite{Gerard et~al.}{2009}]{gerard_study_2009}
Gerard, P., A.~Leonard, J.-P. Masekanya, R.~Charlier, and
  F.~Collin\leavevmode\nopagebreak\newline 2009.
\newblock Study of the soil-atmosphere moisture exchanges through convective
  drying tests in non-isothermal conditions.
\newblock P.~~24.

\bibitem[\protect\astroncite{Geuzaine and Remacle}{2009}]{geuzaine_gmsh_2009}
Geuzaine, C. and J.-F. Remacle\leavevmode\nopagebreak\newline 2009.
\newblock Gmsh: {A} 3-{D} finite element mesh generator with built-in pre- and
  post-processing facilities: {THE} {GMSH} {PAPER}.
\newblock {\em International Journal for Numerical Methods in Engineering},
  79(11):1309--1331.

\bibitem[\protect\astroncite{Ghommem et~al.}{2016}]{ghommem_complexity_2016}
Ghommem, M., E.~Gildin, and M.~Ghasemi\leavevmode\nopagebreak\newline 2016.
\newblock Complexity {Reduction} of {Multiphase} {Flows} in {Heterogeneous}
  {Porous} {Media}.
\newblock {\em SPE Journal}, 21(01):144--151.

\bibitem[\protect\astroncite{Gildin et~al.}{2013}]{gildin_nonlinear_2013}
Gildin, E., M.~Ghasemi, A.~Romanovskay, and
  Y.~Efendiev\leavevmode\nopagebreak\newline 2013.
\newblock Nonlinear {Complexity} {Reduction} for {Fast} {Simulation} of {Flow}
  in {Heterogeneous} {Porous} {Media}.
\newblock In {\em All {Days}}, Pp.~ SPE--163618--MS, The Woodlands, Texas, USA.
  SPE.

\bibitem[\protect\astroncite{Hamade}{2013}]{hamade_geotechnical_2013}
Hamade, T.\leavevmode\nopagebreak\newline 2013.
\newblock {\em Geotechnical {Design} of {Tailings} {Dams} - {A} {Stochastic}
  {Analysis} {Approach}}.
\newblock PhD thesis, McGill University.

\bibitem[\protect\astroncite{Heshmati~R.
  et~al.}{2020}]{heshmati_r_prediction_2020}
Heshmati~R., A.~A., H.~Salehzadeh, and
  M.~Shahidi\leavevmode\nopagebreak\newline 2020.
\newblock Prediction of the {Void} {Ratio} {Parameter} in {Mineral} {Tailings}
  {Using} {Gene} {Expression} {Programming}.
\newblock {\em Advances in Civil Engineering}, 2020:1--12.

\bibitem[\protect\astroncite{Hesthaven et~al.}{2016}]{hesthaven_certified_2016}
Hesthaven, J.~S., G.~Rozza, and B.~Stamm\leavevmode\nopagebreak\newline 2016.
\newblock {\em Certified {Reduced} {Basis} {Methods} for {Parametrized}
  {Partial} {Differential} {Equations}}, {SpringerBriefs} in {Mathematics}.
\newblock Cham: Springer International Publishing.

\bibitem[\protect\astroncite{Hoang et~al.}{2018}]{HOANG201896}
Hoang, K.~C., T.-Y. Kim, and J.-H. Song\leavevmode\nopagebreak\newline 2018.
\newblock Fast and accurate two-field reduced basis approximation for
  parametrized thermoelasticity problems.
\newblock {\em Finite Elements in Analysis and Design}, 141:96--118.

\bibitem[\protect\astroncite{Hui et~al.}{2018}]{hui_real-time_2018}
Hui, S.~R., L.~Charlebois, and C.~Sun\leavevmode\nopagebreak\newline 2018.
\newblock Real-time monitoring for structural health, public safety, and risk
  management of mine tailings dams.
\newblock {\em Canadian Journal of Earth Sciences}, 55(3):221--229.

\bibitem[\protect\astroncite{Knutsson et~al.}{2016}]{knutsson_slope_2016}
Knutsson, R., A.~Bjelkevik, and S.~Knutsson\leavevmode\nopagebreak\newline
  2016.
\newblock Slope stability in landform design.
\newblock In {\em Proceedings of the 11th {International} {Conference} on
  {Mine} {Closure}}, A.~Fourie and M.~Tibbett, eds., Pp.~ 89--98. Australian
  Centre for Geomechanics.
\newblock event-place: Perth.

\bibitem[\protect\astroncite{Kossoff et~al.}{2014}]{kossoff_mine_2014}
Kossoff, D., W.~Dubbin, M.~Alfredsson, S.~Edwards, M.~Macklin, and
  K.~Hudson-Edwards\leavevmode\nopagebreak\newline 2014.
\newblock Mine tailings dams: {Characteristics}, failure, environmental
  impacts, and remediation.
\newblock {\em Applied Geochemistry}, 51:229--245.

\bibitem[\protect\astroncite{Larion et~al.}{2020}]{larion_2020}
Larion, Y., S.~Zlotnik, T.~Massart, and P.~Díez\leavevmode\nopagebreak\newline
  2020.
\newblock Building a certified reduced basis for coupled
  thermo-hydro-mechanical systems with goal-oriented error estimation.
\newblock {\em Computational Mechanics}, 66.

\bibitem[\protect\astroncite{Lu and Cui}{2006}]{lu2006three}
Lu, M.-l. and L.~Cui\leavevmode\nopagebreak\newline 2006.
\newblock Three-dimensional seepage analysis for complex topographical tailings
  dam.
\newblock {\em Yantu Lixue(Rock and Soil Mechanics)}, 27(7):1176--1180.

\bibitem[\protect\astroncite{Lyu et~al.}{2019}]{lyu_comprehensive_2019}
Lyu, Z., J.~Chai, Z.~Xu, Y.~Qin, and J.~Cao\leavevmode\nopagebreak\newline
  2019.
\newblock A {Comprehensive} {Review} on {Reasons} for {Tailings} {Dam}
  {Failures} {Based} on {Case} {History}.
\newblock {\em Advances in Civil Engineering}, 2019:1--18.

\bibitem[\protect\astroncite{Maday and Ronquist}{2004}]{maday_reduced_2004}
Maday, Y. and E.~M. Ronquist\leavevmode\nopagebreak\newline 2004.
\newblock The {Reduced} {Basis} {Element} {Method}: {Application} to a
  {Thermal} {Fin} {Problem}.
\newblock {\em SIAM Journal on Scientific Computing}, 26(1):240--258.

\bibitem[\protect\astroncite{Maday and
  Rønquist}{2002}]{maday_reduced-basis_2002}
Maday, Y. and E.~M. Rønquist\leavevmode\nopagebreak\newline 2002.
\newblock A reduced-basis element method.
\newblock {\em Comptes Rendus Mathematique}, 335(2):195--200.

\bibitem[\protect\astroncite{Martin and
  McRoberts}{1999}]{martin_considerations_1999}
Martin, T.~E. and E.~C. McRoberts\leavevmode\nopagebreak\newline 1999.
\newblock Some considerations in the stability analysis of upstream tailings
  dams.
\newblock P.~~18.

\bibitem[\protect\astroncite{Morton}{2021}]{morton_use_2021}
Morton, K.~L.\leavevmode\nopagebreak\newline 2021.
\newblock The {Use} of {Accurate} {Pore} {Pressure} {Monitoring} for {Risk}
  {Reduction} in {Tailings} {Dams}.
\newblock {\em Mine Water and the Environment}, 40(1):42--49.

\bibitem[\protect\astroncite{Negri et~al.}{2015}]{negri_efficient_2015}
Negri, F., A.~Manzoni, and D.~Amsallem\leavevmode\nopagebreak\newline 2015.
\newblock Efficient model reduction of parametrized systems by matrix discrete
  empirical interpolation.
\newblock {\em Journal of Computational Physics}, 303:431--454.

\bibitem[\protect\astroncite{Nuth and Laloui}{2008}]{nuth_effective_2008}
Nuth, M. and L.~Laloui\leavevmode\nopagebreak\newline 2008.
\newblock Effective stress concept in unsaturated soils: {Clarification} and
  validation of a unified framework.
\newblock {\em International Journal for Numerical and Analytical Methods in
  Geomechanics}, 32(7):771--801.

\bibitem[\protect\astroncite{Ortega‐Gelabert
  et~al.}{2020}]{ortegagelabert_fast_2020}
Ortega‐Gelabert, O., S.~Zlotnik, J.~C. Afonso, and
  P.~Díez\leavevmode\nopagebreak\newline 2020.
\newblock Fast {Stokes} {Flow} {Simulations} for {Geophysical}‐{Geodynamic} 
  {Inverse} {Problems} and {Sensitivity} {Analyses} {Based} {On} {Reduced}
  {Order} {Modeling}.
\newblock {\em Journal of Geophysical Research: Solid Earth}, 125(3):25.

\bibitem[\protect\astroncite{Pinyol et~al.}{2008}]{pinyol_rapid_2008}
Pinyol, N.~M., E.~E. Alonso, and S.~Olivella\leavevmode\nopagebreak\newline
  2008.
\newblock Rapid drawdown in slopes and embankments.
\newblock {\em Water Resources Research}, 44(5).

\bibitem[\protect\astroncite{Qiu and Sego}{2001}]{qiu_laboratory_2001}
Qiu, Y.~J. and D.~C. Sego\leavevmode\nopagebreak\newline 2001.
\newblock Laboratory properties of mine tailings.
\newblock {\em Canadian Geotechnical Journal}, 38(1):183--190.

\bibitem[\protect\astroncite{Quarteroni et~al.}{2016}]{quarteroni_reduced_2016}
Quarteroni, A., A.~Manzoni, and F.~Negri\leavevmode\nopagebreak\newline 2016.
\newblock {\em Reduced {Basis} {Methods} for {Partial} {Differential}
  {Equations}}, volume~92 of {\em {UNITEXT}}.
\newblock Cham: Springer International Publishing.

\bibitem[\protect\astroncite{Quarteroni and
  Rozza}{2014}]{quarteroni_reduced_2014}
Quarteroni, A. and G.~Rozza, eds.\leavevmode\nopagebreak\newline 2014.
\newblock {\em Reduced {Order} {Methods} for {Modeling} and {Computational}
  {Reduction}}.
\newblock Cham: Springer International Publishing.

\bibitem[\protect\astroncite{Rozza et~al.}{2007}]{rozza_reduced_2007}
Rozza, G., D.~B.~P. Huynh, and A.~T. Patera\leavevmode\nopagebreak\newline
  2007.
\newblock Reduced basis approximation and a posteriori error estimation for
  affinely parametrized elliptic coercive partial differential equations.
\newblock {\em Archives of Computational Methods in Engineering}, 15(3):1--47.

\bibitem[\protect\astroncite{Saad and Mitri}{2011}]{saad_hydromechanical_2011}
Saad, B. and H.~Mitri\leavevmode\nopagebreak\newline 2011.
\newblock Hydromechanical {Analysis} of {Upstream} {Tailings} {Disposal}
  {Facilities}.
\newblock {\em Journal of Geotechnical and Geoenvironmental Engineering},
  137(1):27--42.

\bibitem[\protect\astroncite{Santo and
  Manzoni}{2019}]{santo_hyper-reduced_2019}
Santo, N.~D. and A.~Manzoni\leavevmode\nopagebreak\newline 2019.
\newblock Hyper-reduced order models for parametrized unsteady
  {Navier}-{Stokes} equations on domains with variable shape.
\newblock {\em Advances in Computational Mathematics}, 45(5-6):2463--2501.

\bibitem[\protect\astroncite{Szostak et~al.}{2003}]{szostak_use_2003}
Szostak, A., A.~Chrzanowski, and M.~Massiéra\leavevmode\nopagebreak\newline
  2003.
\newblock Use of geodetic monitoring measurements in solving geomechanical
  problems in structural and mining engineering.
\newblock P.~~9, Santorini, Greece.

\bibitem[\protect\astroncite{van Doren
  et~al.}{2006}]{van_doren_reduced-order_2006}
van Doren, J. F.~M., R.~Markovinović, and J.-D.
  Jansen\leavevmode\nopagebreak\newline 2006.
\newblock Reduced-order optimal control of water flooding using proper
  orthogonal decomposition.
\newblock {\em Computational Geosciences}, 10(1):137--158.

\bibitem[\protect\astroncite{van
  Genuchten}{1980}]{van_genuchten_closed-form_1980}
van Genuchten, M.~T.\leavevmode\nopagebreak\newline 1980.
\newblock A {Closed}-form {Equation} for {Predicting} the {Hydraulic}
  {Conductivity} of {Unsaturated} {Soils}.
\newblock {\em Soil Science Society of America Journal}, 44(5):892--898.

\bibitem[\protect\astroncite{Vanden~Berghe
  et~al.}{2011}]{vanden_berghe_geotechnical_2011}
Vanden~Berghe, J.-F., J.-C. Ballard, J.-F. Wintgens, and
  B.~List\leavevmode\nopagebreak\newline 2011.
\newblock Geotechnical {Risks} {Related} to {Tailings} {Dam} {Operations}.
\newblock In {\em Tailings and {Mine} {Waste} {Conference} (2011 : {Vancouver},
  {B}.{C}.)}, P.~~11.

\bibitem[\protect\astroncite{Vermeulen et~al.}{2004}]{vermeulen_reduced_2004}
Vermeulen, P., A.~Heemink, and C.~Te~Stroet\leavevmode\nopagebreak\newline
  2004.
\newblock Reduced models for linear groundwater flow models using empirical
  orthogonal functions.
\newblock {\em Advances in Water Resources}, 27(1):57--69.

\bibitem[\protect\astroncite{Villavicencio
  et~al.}{2011}]{villavicencio_estimation_2011}
Villavicencio, A.~G., P.~Breul, C.~Bacconnet, D.~Boissier, and A.~R.
  Espinace\leavevmode\nopagebreak\newline 2011.
\newblock Estimation of the {Variability} of {Tailings} {Dams} {Properties} in
  {Order} to {Perform} {Probabilistic} {Assessment}.
\newblock {\em Geotechnical and Geological Engineering}, 29(6):1073--1084.

\bibitem[\protect\astroncite{Zhang et~al.}{2020}]{zhang_numerical_2020}
Zhang, C., J.~Chai, J.~Cao, Z.~Xu, Y.~Qin, and
  Z.~Lv\leavevmode\nopagebreak\newline 2020.
\newblock Numerical {Simulation} of {Seepage} and {Stability} of {Tailings}
  {Dams}: {A} {Case} {Study} in {Lixi}, {China}.
\newblock {\em Water}, 12(3):742.

\end{thebibliography}

\end{document}